\documentclass[preprint2]{aastex}

\def\mc{{WHL~J085910.0+294957}}

\shorttitle{Faint Companions of Bright Galaxies in \mc}

\shortauthors{Lee et al.}
\def\simlt{\lower.5ex\hbox{$\; \buildrel < \over \sim \;$}}
\def\simgt{\lower.5ex\hbox{$\; \buildrel > \over \sim \;$}}

\begin{document}

\title{MEASURABLE RELATIONSHIP BETWEEN BRIGHT GALAXIES AND THEIR FAINT COMPANIONS IN \mc, A GALAXY CLUSTER AT {\it z} = 0.30: VESTIGES OF INFALLEN GROUPS?}

\author{Joon Hyeop Lee$^{1,2}$, Hye-Ran Lee$^{1,2}$, Minjin Kim$^{1,2,3\ \dag}$, Kwang-Il Seon$^{1,2}$, Sang Chul Kim$^{1,2}$, Soung-Chul Yang$^{1,3\ \dag}$, Chang Hee Ree$^1$, Jong Chul Lee$^1$, Hyunjin Jeong$^{1,2}$, Jongwan Ko$^1$, Changsu Choi$^{4}$}
\affil{$^1$ Korea Astronomy and Space Science Institute, Daejeon 305-348, Republic of Korea\\
$^2$ Korea University of Science and Technology, Daejeon 305-350, Republic of Korea\\
$^3$ The Observatories of the Carnegie Institution for Science, 813 Santa Barbara Street, Pasadena, CA 91101, USA\\
$^4$ CEOU/Astronomy Program, Department of Physics and Astronomy, Seoul National University, Seoul 151-742, Republic of Korea\\
$\dag$ KASI-Carnegie Fellowship.}

\email{jhl@kasi.re.kr}

\begin{abstract}
The properties of satellite galaxies are closely related to their host galaxies in galaxy groups. In cluster environments, on the other hand, the interaction between close neighbors is known to be limited. Our goal is to examine the relationships between host and satellite galaxies in the harsh environment of a galaxy cluster. To achieve this goal, we study a galaxy cluster {\mc} at $z=0.30$ using deep images obtained with CQUEAN CCD camera mounted on the 2.1-m Otto Struve telescope. After member selection based on the scaling relations of photometric and structural parameters, we investigate the relationship between bright ($M_i \le -18$) galaxies and their faint ($-18<M_i \le -15$) companions. The weighted mean color of faint companion galaxies shows no significant dependence ($<1\sigma$ to Bootstrap uncertainties) on cluster-centric distance and local luminosity density as well as the luminosity and concentration of an adjacent bright galaxy. However, the weighted mean color shows marginal dependence ($\sim2.2\sigma$) on the color of an adjacent bright galaxy, when the sample is limited to bright galaxies with at least 2 faint companions. By using a permutation test, we confirm that the correlation in color between bright galaxies and their faint companions in this cluster is statistically significant with a confidence level of $98.7\%$. The statistical significance increases if we additionally remove non-members using the SDSS photometric redshift information ($\sim2.6\sigma$ and $99.3\%$). Our results suggest three possible scenarios: (1) vestiges of infallen groups, (2) dwarf capturing, and (3) tidal tearing of bright galaxies.
\end{abstract}

\keywords{galaxies: clusters: individual (\mc) --- galaxies: dwarf --- galaxies: elliptical and lenticular, cD --- galaxies: evolution --- galaxies: formation}

\section{INTRODUCTION}

Today, it is widely accepted that the properties of galaxies are significantly affected by their environments. Galaxies in high-density environments tend to have red colors, early-type morphologies, high masses, poor gas reservoirs, and low star formation rates \citep[e.g.,][]{dre80,kau04,bal06,par07,pog08,lee10}, although such dependence of galaxy properties on environments moderately varies according to the environmental scales \citep{bla07} or the definition of environmental parameters \citep{mul12,haa12}. 
It has been also revealed that such trends of environmental effects vary as a function of redshift \citep[e.g.,][]{but84,sco13}. For example, the locally-observed relationship between star formation and local density is already established at $z\sim0.5$ \citep{got03}, but it appears to be reversed at $z\sim1$, which means that galaxies at high redshifts tend to form stars actively at high-density environments \citep{elb07,coo08,pop11}.

Galaxy clusters are the highest density environments in the Universe, which are thought to accelerate the evolution of galaxies by various mechanisms. In such high-density environments, member galaxies frequently experience galaxy-galaxy close encounters, which can cause galaxy harassment \citep{moo96,moo99} and galaxy-galaxy hydrodynamic interaction \citep{par09}. The deep gravitational potentials of galaxy clusters may also often give rise to the interaction between galaxies and the cluster potential itself \citep{mer84,gne03}. Moreover, since galaxy clusters typically have a large amount of hot gas with high pressure, the internal gas contents of galaxies can be strongly affected by the mechanisms such as ram pressure stripping \citep{gun72,qui00} and strangulation \citep{lar80,bek02}.
On the other hand, in less dense environments like galaxy groups or fields, the tidal interactions between close neighbors or interactions with satellite galaxies also become an important driver for the galaxy evolution:
\citet{byr90} and many other following studies have shown that the physical properties of galaxies (morphology, color, gas and dust contents, star formation, and AGN activities, and so on) are significantly influenced by the tidal interactions between individual galaxies \citep[e.g.,][]{her06,per06,coz07,bes12,lee12,pat13}.

In a small galaxy group consisting of a bright host galaxy and its satellites, the collective properties of the satellite galaxies are expected to affect the properties of the host galaxy through the accretion of satellites onto the massive galaxy \citep{pau13}. As well as such merger events, gas looting by hydrodynamic interaction is another channel how host and satellite galaxies affect each other: typically massive host galaxies tend to deprive their satellites of gas \citep[a good example is the Milky Way Galaxy and Large Magellanic Cloud; e.g.,][]{mas05}. It is also known that the tidal stripping efficiency of satellites depends on the morphology of their host galaxy as well as their own morphologies \citep{cha13}.
Satellite galaxies in a group seem to suffer transformation in color and morphology possibly through strangulation, disk fading, galaxy mergers or close tidal encounters \citep{geo13}.
As a result, the luminosity function and spatial distribution of satellites depend on the properties of their host galaxy \citep{lar11} and the properties of satellite galaxies are known to be sensitive to the group properties such as halo mass, local density and dynamical state \citep{pen12,car13,woo13}.
The early-type fraction of satellites is found to be significantly higher in a halo with an early-type host galaxy than in a halo with a late-type host galaxy with the same mass \citep[so-called `galactic conformity';][]{wei06,ann08}. More recently, \citet{phi13} argued that the satellites of bright isolated galaxies show difference in their star formation activities according to the host properties, in the sense that $30\%$ of satellites around quiescent host galaxies suffer star formation quenching, while $0\%$ around star-forming host galaxies do.

From these environmental effects in different - large and small - scales, one question rises: how do the effects in different scales work in a high-density environment such as a galaxy cluster? In other words, in a galaxy cluster where the large scale environmental effects are strong, how significant are the effects of small scale interactions?
It was previously pointed out that the encounter time between cluster galaxies is too short, due to the high velocity dispersion of a galaxy cluster, to produce tidal energy enough to significantly affect galaxy structure \citep{mer84,byr90,bos06}.
From observational data, \citet{par09} showed that the properties of galaxies suddenly start to depend on the cluster-centric distance at fixed neighbor environment with a characteristic scale of $1-3\;\times$ virial radius, which implies that the effect of direct galaxy-galaxy tidal interactions is less significant in a galaxy cluster, although they argued that the galaxy-galaxy hydrodynamic interactions still play an important role even in a cluster environment.
The current consensus is that the role of tidal interactions between galaxies is not crucial to determine the properties of galaxies in a galaxy cluster.
However, the galaxy sample in most previous studies are limited to bright galaxies, and faint dwarf galaxies are hardly considered; for example, \citet{par09} used a sample of galaxies with $M_r\le -17$. Thus, currently it is not certain whether the properties of bright cluster galaxies are related to their faint companions or not. In other words, the close relationship between host and satellite galaxies shown in a group scale is yet to be investigated in galaxy cluster environments.
Our goal is to statistically address the relationships between bright cluster galaxies and their \emph{faint} companions (i.e., their possible satellites).

In this paper, to achieve our goal, we carry out a case study of {\mc}\footnote{This cluster is also called as CMBCG J134.79176+29.83268 or MaxBCG J134.79176+29.83268.}, a galaxy cluster at $z=0.30$, using two-band deep photometric data.
The outline of this paper is as follows. Section 2 describes our observations, data reduction and photometry. The analysis methods are shown in Section 3, including cluster member selection and parameter definitions. The results are presented in Section 4 and their implication is discussed in Section 5. Section 6 gives the conclusion. Throughout this paper, we adopt the cosmological parameters: $h=0.7$, $\Omega_{\Lambda}=0.7$, and $\Omega_{M}=0.3$.

\section{OBSERVATIONS AND DATA REDUCTION}

\subsection{Target Information}

Our target cluster {\mc} was identified by using the Sloan Digital Sky Survey \citep[SDSS;][]{yor00} data in \citet{koe07} and \citet{hao10}. They detected optically rich galaxy clusters, based on the fact that rich clusters typically show tight red sequences and many of them have Brightest Cluster Galaxies (BCGs) in their centers (maxBCG red-sequence method; see those papers for more details). Among the tens of thousand galaxy clusters listed in \citet{koe07} and \citet{hao10}, we first selected target candidates so that the telescope field-of-view (see Section~\ref{observe}) covers at least 1 Mpc diameter of the targets. After that, we chose {\mc} at $z=0.2998$ as the final target cluster, which shows very obvious clustering of bright galaxies in the visual inspection. The spectroscopic redshift of the BCG is available ($z=0.2656$), but the representative redshift value was estimated from the photometric redshifts of the 35 member galaxies. The scaled richness\footnote{\citet{koe07} defined the \emph{
scaled} richness as ``the number of E/S0 member galaxies brighter than $0.4L^*$ and within $R_{200}$ of the cluster center, $R_{200}$ being the radius within which the density of galaxies with $-24\le M_r\le-16$ is 200 times the mean density of such galaxies''. Note that this value is larger than the \emph{measured} richness, 35, which is the number of members found in their cluster-finding algorithm.} is 41.
The right ascension and declination (J2000) of the cluster are $8^h59^m10^s$ and $29^{\circ}49'57''$ respectively, which was determined from the coordinate of the BCG \citep{koe07}.

\subsection{Observations}\label{observe}

\begin{figure}[t]
\plotone{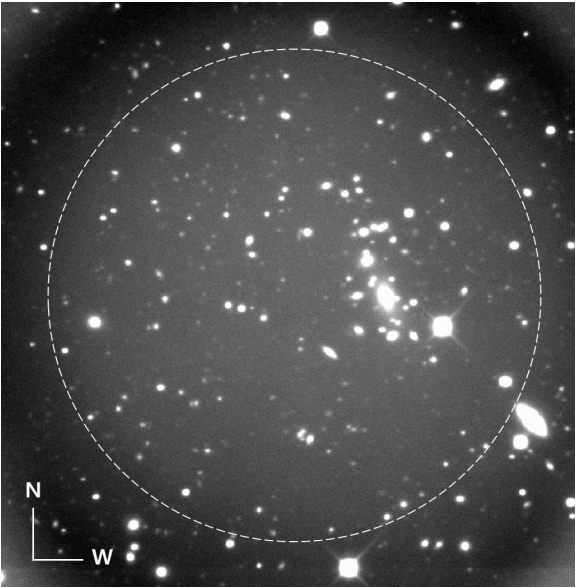}
\caption{ The stacked $i$-band image of {\mc}, observed using the Otto Struve 2.1-m telescope and CQUEAN CCD camera. The field of view is $4.7'\times4.7'$, and the dashed circle displays the analysis area with a radius of 400 pixels ($\sim 110''$).
\label{obs}}
\end{figure}

The observations were carried out on 9 - 10 February 2012, using the Camera for QUasars in EArly uNiverse \citep[CQUEAN;][]{par12} mounted on the 2.1-m Otto Struve Telescope in the McDonald Observatory, USA.
CQUEAN is a CCD camera with 1024 by 1024 pixels and its field of view is $4.7'\times 4.7'$ when mounted on the 2.1-m telescope, which corresponds to 0.276 arcsecond per pixel. CQUEAN is designed to efficiently detect long-wavelength visible and near-infrared light. During the two clear nights, total exposures are 13,700 and 17,460 seconds in the $r$ and $i$ bands, respectively. The stacked $i$-band image of {\mc} is displayed in Figure~\ref{obs}. The typical seeing size ranges from $1.5''$ to $2.0''$, which corresponds to the physical scale of $6.7 - 8.9$ kpc at $z=0.3$.

In the observed images, the central BCG is clearly identified and several bright galaxies are found to be linearly aligned around the BCG. The existence of a distinct BCG indicates that this cluster started to be assembled long time ago \citep[e.g.,][]{del07}. However, the linear structure of bright galaxies implies that this cluster is not in a dynamically relaxed state currently.

\subsection{Data Reduction and Photometry}\label{phot}

\begin{figure}[t]
\plotone{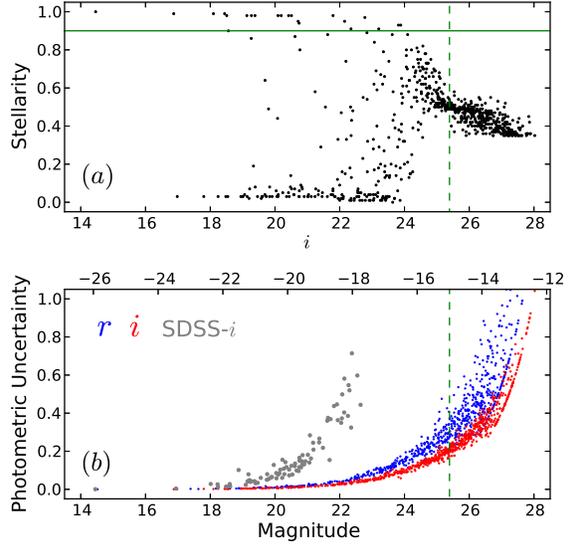}
\caption{ (a) Stellarity versus $i$-band apparent magnitude. The horizontal line shows the criterion dividing the objects into stars and galaxies: stellarity $= 0.9$.
(b) Photometric uncertainty along magnitude in the $r$ (blue dots) and $i$ (red dots) bands. The grey dots show the $i$-band SDSS data for the comparison of observational depth. The lower axis ticks show apparent magnitudes, while the upper axis ticks display absolute magnitudes on the assumption that all objects are at $z=0.3$ (throughout Figures~\ref{error} -- \ref{glf}). The vertical dashed lines indicate $i=25.39$ (corresponding to $M_i=-15$ at $z=0.3$), which is the magnitude limit in our analysis.
\label{error}}
\end{figure}

The observed images were processed using the standard \emph{IRAF} \citep[Image Reduction and Analysis Facility;][]{tod86} packages: pre-processing, aligning and image combining. The \emph{Source Extractor} \citep{ber96} was used for our photometry of galaxies, in \emph{dual mode} with $i$-band reference.
Standard calibration and astrometry were conducted by comparing our catalog with the SDSS Data Release 7 \citep[SDSS DR7;][]{aba09}. 
Figure~\ref{error}(a) shows the distribution of \emph{stellarities} along magnitude provided by the \emph{Source Extractor}, in which we select the objects with stellarity $\le 0.9$ as galaxies. Figure~\ref{error}(b) displays the photometric error versus magnitude of the selected galaxies in our images \footnote{The distance modulus adopted for this galaxy cluster is $m - M = 40.39$.}. All magnitudes in this paper are in the AB system.
To secure reasonable signal-to-noise ratios in detection, we use objects brighter than $i=25.39$ ($M_i=-15$ at $z=0.3$). At $i=25.39$, the $i$-band magnitude uncertainty is about 0.2, which approximately corresponds to the signal-to-noise ratio of 5.
The depth of our observation is deeper by $\sim4$ mag than that of the SDSS; our observation newly added faint objects with $21\lesssim i \lesssim25$, which roughly corresponds to $0.01 L_{*}\lesssim L \lesssim 0.4 L_{*}$ at $z=0.3$ \citep[when the characteristic magnitude $i_{*}\approx20$ is supposed;][and Section \ref{memsel} in this paper]{har07}.

The number of objects satisfying $i\le 25.39$ and stellarity $\le 0.9$ is 901.
However, the outer parts of CQUEAN images suffer from vignetting as shown in Figure~\ref{obs}. Since the area farther than 400 pixels from the image center shows significant decrease in the mean background level (larger than $3\sigma$ of local background fluctuation), we use only the objects within 400 pixels from the image center (the \emph{analysis area} denoted with a dashed circle in Figure~\ref{obs}). The final catalog of galaxies includes 402 objects.
The correction for foreground reddening has been carried out using the reddening map of \citet{sch98} and the reddening law of \citet{car89}: $E(B-V) = 0.029$, $A_r = 0.080$ and $A_i = 0.061$. K-correction has not been applied to the cluster galaxies.

\section{ANALYSIS}\label{anal}

\subsection{Cluster Member Selection}\label{memsel}

The 402 galaxies selected in Section~\ref{phot} are not necessarily the members of the galaxy cluster {\mc}, but this sample may include many foreground and background objects. If there exist some very strong intrinsic trends in the genuine cluster member galaxies, we may be able to catch signals for the trends even though we do not remove contamination. For example, tight red sequences are easily found in most rich clusters even without member selection. However, since the background and foreground contamination would considerably decrease the statistical significance of any signal, we need to reduce such contamination as well as possible to get clearer results.
The best way to remove foreground and background contamination is to use their redshift information. Unfortunately, however, we do not have the spectroscopic data for most galaxies in our sample. Moreover, even if there were chances of multi-object spectroscopic observations, it would be very difficult to secure the spectra of many faint dwarf galaxies in this galaxy cluster due to the observational limitation.

There are some alternative methods for the membership selection of cluster galaxies such as control field subtraction \citep[e.g.,][]{pao01} and red sequence selection \citep[e.g.,][]{dep98}.
The control field subtraction is a statistical method typically used for deriving cluster galaxy luminosity functions. However, this is not suitable for our purpose, because this method removes contamination not individually but statistically and collectively (in other words, we can not distinguish whether a given individual galaxy is a member or not). On the other hand, the red sequence selection may be used for our study in the sense that it removes contamination individually. However, this method also has an obvious limitation, because it selects only red galaxies and misses possible cluster members with blue colors. 

\begin{figure}[t]
\plotone{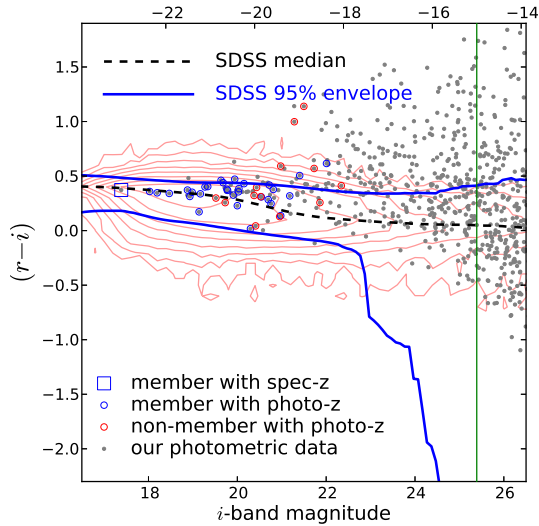}
\caption{ Cluster member selection using the color-magnitude relation (CMR). The dots are the galaxies in our photometry. The contours show the logarithmic number density distribution of the Sloan Digital Sky Survey (SDSS) galaxies at $0.001<z<0.300$, k-corrected to $z=0.3$. The thick lines connect the median colors (dashed) and the $95\%$ envelopes (solid) of the SDSS galaxies. The vertical line denotes the magnitude limit in our analysis: $M_i=-15$. For comparison, the objects with known spectroscopic (open square) or photometric (open circles) redshifts are denoted, distinguished between members (blue; at $0.2\le$ photo-z $\le0.4$) and non-members (red).
\label{cmr}}
\end{figure}

\begin{figure}[t]
\plotone{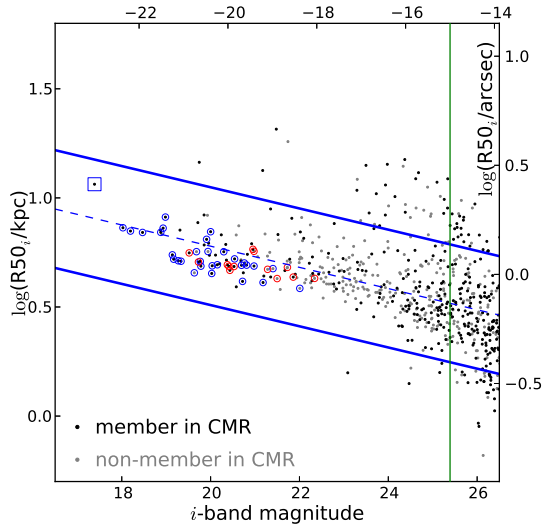}
\caption{ Cluster member selection using the size-magnitude relation (SMR). The black dots are the galaxies selected in Figure~\ref{cmr} and the grey dots are the rejected ones. The thin and dashed line is the result of the iterative linear least squares fit, while the thick lines display the member selection criteria. The vertical line denotes the magnitude limit in our analysis: $M_i=-15$. The Y-axis ticks in the right side are the angular sizes, while those in the left side show the physical sizes when supposing that the objects are at $z=0.3$. The open symbols show the known membership with spectroscopic or photometric redshifts like in Figure~\ref{cmr}.
\label{size}}
\end{figure}

\begin{figure}[t]
\plotone{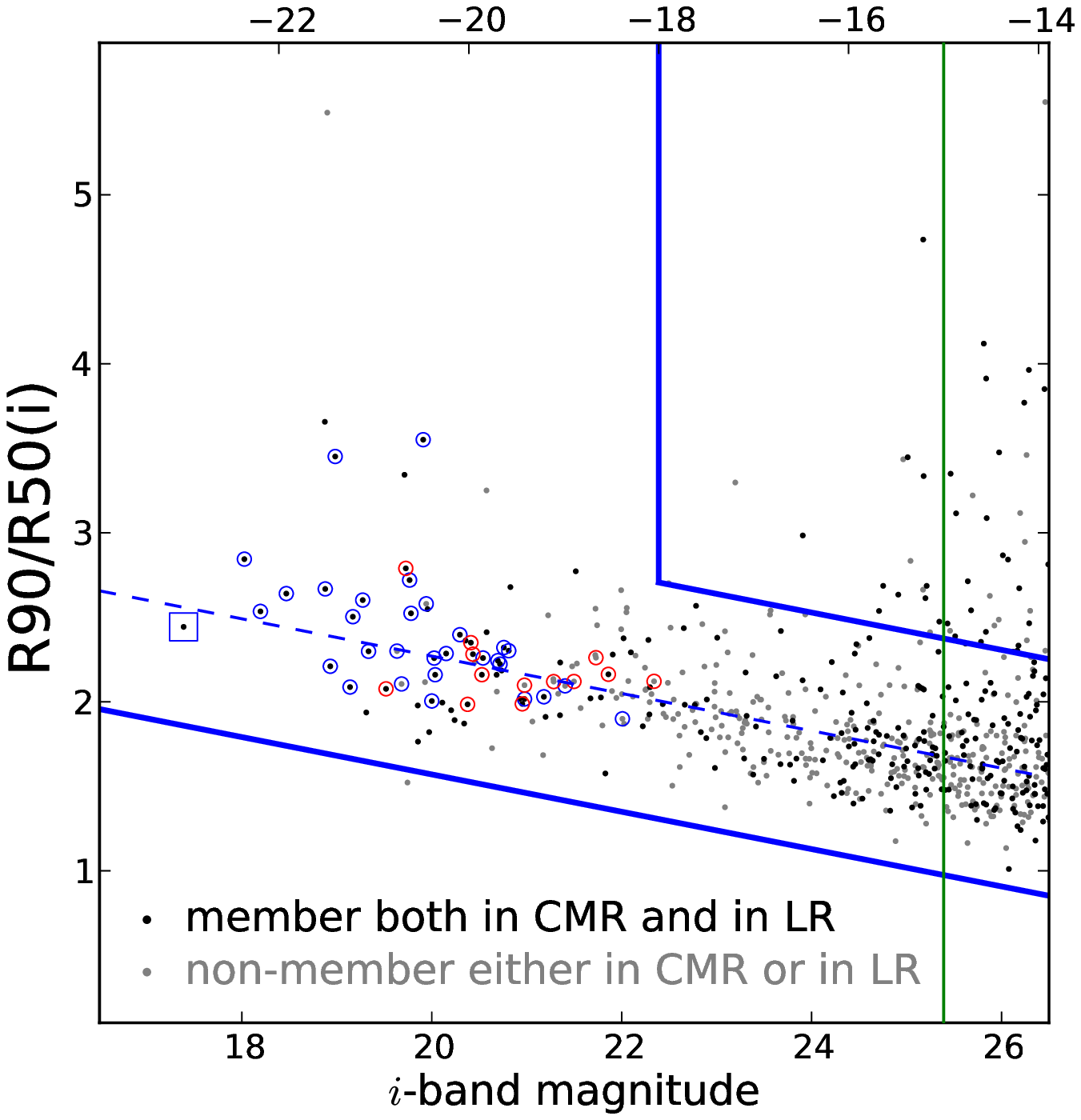}
\caption{ Cluster member selection using the concentration-magnitude relation (ConMR). The black dots are the galaxies selected in Figures~\ref{cmr} and \ref{size}, while the grey dots are the rejected ones. The thin and dashed line is the result of the iterative linear least squares fit, while the thick lines display the member selection criteria. The thin and vertical line denotes the magnitude limit in our analysis: $M_i=-15$. The open symbols show the known membership with spectroscopic or photometric redshifts like in Figure~\ref{cmr}.
\label{con}}
\end{figure}

In this paper, we extend the concept of red sequence selection method to choose individual member galaxies better. That is, we adopt three scaling relations of galaxies: color-magnitude relation (CMR), size-magnitude relation (SMR), and concentration-magnitude relation (ConMR). 

In the CMR method, we select cluster members based on the color distribution \emph{expected} at $z=0.3$, which was estimated using the SDSS data as shown in Figure~\ref{cmr}, instead of cutting a narrow range of colors as is done in the red sequence selection.
To derive the $z=0.3$ expected color distribution, we first corrected the systematic differences between the magnitudes and colors in this paper and those in the SDSS (mainly due to the difference in aperture definition), which were measured using 6 bright ($i<18$) galaxies found both in the SDSS and in our catalog to be:
\begin{eqnarray}
i_{\textrm{\scriptsize TP}} & = & i_{\textrm{\scriptsize SDSS}} + 0.0787 \\
(r-i)_{\textrm{\scriptsize TP}} & = & (r-i)_{\textrm{\scriptsize SDSS}} - 0.1732,
\end{eqnarray}
where $i_{\textrm{\scriptsize TP}}$ and $(r-i)_{\textrm{\scriptsize TP}}$ denote the $i$-band magnitude and $r-i$ color in this paper (TP), respectively. The values of 0.0787 and $-0.1732$ are the medians for the six galaxies, where the sample interquartile ranges are 0.0123 and 0.0109 for $\Delta i$ and $\Delta (r-i)$, respectively.
After that, the expected colors of SDSS galaxies with spectroscopic redshifts when they \emph{moved to $z=0.3$} were calculated using the method of \citet{bla03}. We defined envelopes including $95\%$ of those SDSS galaxies as shown in Figure~\ref{cmr}, which we regard as the \emph{expected} CMR range at $z=0.3$. With this criterion, 194 of 402 galaxies are selected as cluster members.

The second scaling relation is the SMR \citep[e.g.,][]{tru04,nai10,lee13}. Although the SMR is not tight enough to select cluster members accurately, we can use it to remove objects with obviously unreasonable sizes (i.e., too deviated from the SMR).
This is possible because the angular sizes are the combinations of physical sizes and redshifts. Since the ratio of luminosity distance to angular diameter distance is $(1+z)^2$, the objects at different redshifts are expected to show difference in the relation between their angular sizes and apparent magnitudes.

In this selection, however, it is not easy to directly compare the SMR in our data with that in some other data such as the SDSS, because the size is a parameter sensitive to background estimation and total luminosity definition. Hence, instead of using the SMR itself from the SDSS, we brought only the \emph{dispersion} in the SMR (more exactly, the size dispersion in physical scale at given absolute magnitude) from the SDSS spectroscopic data. With the dispersion value from the SDSS and the SMR estimated using our own data, the process to select the cluster members goes as follows: (i) Derive the linear least squares fit in the size-magnitude diagram using our galaxy sample with $i\le 25.39$. (ii) Build a sub-sample including galaxies within $\pm0.27$ in $\log R50_i$ from the estimated linear relation, where $R50_i$ is the semimajor axis length of ellipse containing $50\%$ of the Petrosian flux in the $i$ band. Here, the dispersion cut 0.27 is the value from the SDSS SMR, which includes $95\%$ of the SDSS 
galaxies. (iii) Using the sub-sample, re-estimate the linear least squares fit. (iv) Repeat (ii) and (iii) one more 
time (resulting in a total of three times iteration for the clipping-and-fitting process).

The finally estimated SMR is:
\begin{equation}
\log (R50_i/\textrm{kpc}) = -0.0486 \times M_i - 0.2126.
\end{equation}
We reject the objects deviated by more than 0.27 in $\log R50_i$ from this relation, as shown in Figure~\ref{size}. As a result, 46 of 402 (or 28 of the CMR-selected 194) objects are rejected from our cluster member sample, because they are too large or too small compared to their magnitudes.

Similarly, we select cluster members in the ConMR, as presented in Figure~\ref{con}. The light concentration itself is a parameter that is hardly affected by redshift, but concentration values are known to strongly depend on absolute magnitudes in the sense that more luminous galaxies tend to be more concentrated on average \citep[e.g.,][]{lee08}. This fact can be used in rough selection of cluster members; for example, \citet{lie12} used the S{\'e}rsic index cut to select faint members of the Virgo cluster, where S{\'e}rsic index is closely related to light concentration \citep[e.g.,][]{gra01}.
Here, we apply a selection process similar to the selection based on the SMR described in the previous paragraphs. The clipping range is selected to be $\pm0.7$ in $R90/R50(i)$, which was derived from the ConMR of the SDSS galaxies with spectroscopic information, including $95\%$ of the SDSS galaxies. $R90/R50(i)$ is the ratio between semimajor axis lengths of ellipses containing $90\%$ and $50\%$ of the Petrosian flux in the $i$ band.

The finally estimated ConMR is:
\begin{equation}
\log (R90/R50(i)) = -0.1105 \times M_i + 0.0166.
\end{equation}
We reject the objects deviated by more than 0.7 in $R90/R50(i)$ from this relation at $M_i>-18$, but we do not reject highly-concentrated galaxies at $M_i\le -18$ because bright early-type galaxies often have very high concentration values \citep[e.g.,][]{ber10}.
As a result, 21 of 402 (or 10 of the CMR-and-SMR-selected 166) objects are rejected from our cluster member sample due to their unusual concentration values.
The final cluster members are selected by applying the three criteria simultaneously, which yields the final sample of 156 galaxies.

Note that the SMR and ConMR selections have a risk at faint end, because the sizes of very faint objects can be mis-measured. Our SMR and ConMR criteria may reject such possibly mis-measured objects. However, if there is a problem in size measurement for a given object, the flux measurement for that object may also have a problem, and thus it may be better to reject such an object for higher reliability in analysis.

\begin{figure}[t]
\plotone{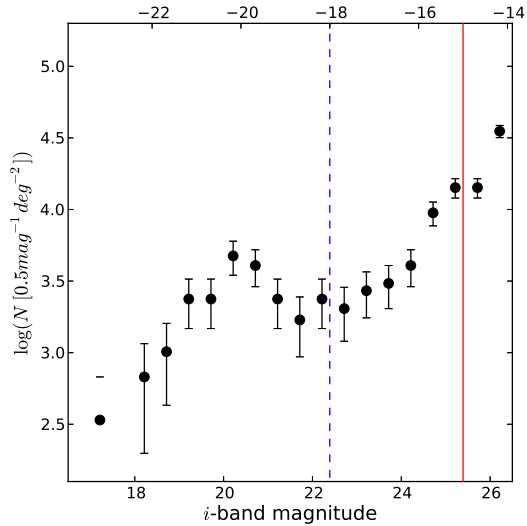}
\caption{ Galaxy luminosity function for the selected members of \mc. The vertical lines show $M_i=-18$ (dashed) and $M_i=-15$ (solid). The bars indicate Poisson errors.
\label{glf}}
\end{figure}

Among the selected cluster members, we define as \emph{bright galaxies} the ones with $M_i\le-18$, which yields 65 \emph{bright galaxies}, and as their \emph{faint companions} (possible satellites) the ones with $M_i>-18$, closer than 100 kpc \footnote{\citet{par09} estimated the typical virial radius of early-type galaxies with $M_r=-19.5$ to be $260\;h^{-1}$ ($\sim 370$) kpc. Our 100 kpc criterion is much smaller than that typical value, which means that \emph{faint companions} can be considerably affected by their adjacent \emph{bright galaxies}. However, it should be always kept in mind that we are looking just projected distances. } to individual \emph{bright galaxies}.
These \emph{bright galaxies} and their \emph{faint companions} do NOT necessarily have genuine host-satellite relationships sharing dark matter halos, to confirm which dynamical information of those objects is required. Thus, we regard those galaxies as \emph{candidates} of genuine host and satellite galaxies.
The magnitude cut of $M_i=-18$ is much fainter than typical characteristic luminosity in a galaxy luminosity function, but the fraction of blue cloud galaxies is known to start increasing significantly at about this magnitude \citep{dep13}. In Figure~\ref{glf}, the bimodal distribution of galaxy luminosity is quite well divided into bright and faint groups by the $M_i=-18$ cut.

\begin{figure}[t]
\plotone{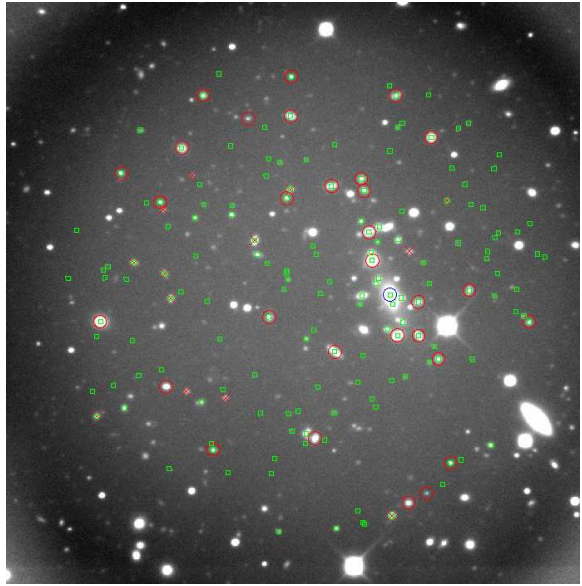}
\caption{ The stacked $i$-band image of {\mc}, with the selected cluster members highlighted (open squares). The spec-z/photo-z members (open blue/red circles) and photo-z non-members (crosses) are also denoted.
\label{obsmem}}
\end{figure}

In addition to our main sample of cluster members described until now, we consider photometric redshifts (hereafter, photo-z's) retrieved from the SDSS, for 43 bright objects within our analysis area. Since the SDSS photo-z's are not available at $i>22$ and the photo-z estimation uncertainties are quite large (up to $\sim0.1$ at $i\sim21$ and even larger at $21<i<22$), the member selection based on only photo-z's is not suitable for our purpose, particularly for the selection of faint members.
However, it is still useful to cross-check the photo-z membership with our selected bright members. Here, we select photo-z members as the objects at $0.2\le$ photo-z $\le0.4$. This seems to be a somewhat generous criterion, but may be appropriate when we consider the large uncertainties in photo-z estimation. This criterion yields 30 photo-z members and 13 photo-z non-members, which are overplotted in Figures \ref{cmr} -- \ref{con}. Note that 5 photo-z members are rejected in our main sample, while 8 photo-z non-members are included. In the subsequent analysis, the sample without any additional notation indicates our main sample. On the other hand, `$+$ photo-z' notation means a sample in which 5 photo-z members are added to the main sample, while `$-$ photo-z' notation indicates a sample in which 8 photo-z non-members are removed from the main sample. Figure~\ref{obsmem} shows the spatial distribution of our selected cluster members, and the photo-z members and non-members.

\subsection{Completeness and Reliability}\label{comrel}

\begin{figure}[t]
\plotone{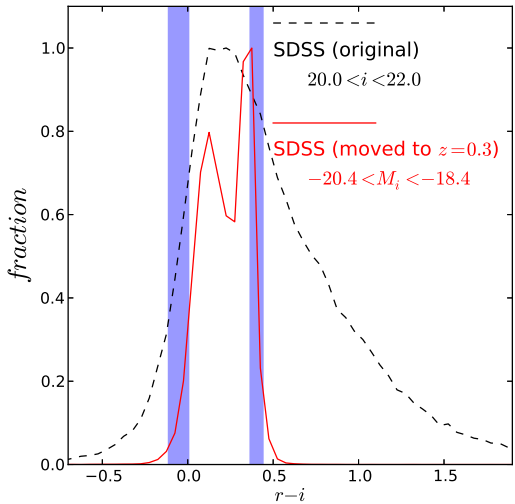}
\caption{ Completeness and reliability test for the CMR selection at $20.0<i<22.0$. The shaded vertical stripes show our member selection criteria using the CMR, which are not lines but stripes because the criteria vary along magnitude. The solid line displays the color distribution of the SDSS galaxies moved to $z=0.3$, while the dashed line shows the apparent color distribution of the SDSS galaxies in the $3^\circ\times3^\circ$ field centered on {\mc} without K-correction. Both distribution curves were normalized to their peak values.
\label{comprel}}
\end{figure}

As mentioned previously, our selection is not perfect, which means that considerable contamination may still remain. Thus, in this section, we estimate the completeness (how many genuine members we selected among all genuine members) and reliability (how many genuine members exist among all selected galaxies) in our cluster member selection.

Before doing it, however, we emphasize that some meaningful tendencies may be found even though we do not remove contamination, if the intrinsic trends are strong enough. The cluster red sequence mentioned in the beginning of Section~\ref{memsel} is a good example.
It is certain that the intrinsic trends will be revealed more obviously if we select genuine cluster members better, which means that poor member selection will typically weaken the observed tendencies by adding random scatters to the intrinsic trends. In other words, if we detect some tendencies that can not be the results of selection bias (will be discussed in Section~\ref{uncer}), then the fact that the member selection was imperfect does not weaken the conclusion; on the contrary, we can assume that better member selection would reveal the tendencies more clearly.

Now, our strategy to estimate the completeness and reliability is as follows:
\begin{enumerate}
 \item Estimate the completeness and reliability of the CMR selection, using the photometric and spectroscopic catalogs of the SDSS galaxies.
 \item Based on the completeness and reliability of the CMR selection, additionally estimate the effects of the SMR and ConMR selections.
\end{enumerate}

For the first step, the idea is to compare our color criterion at given magnitude range with (i) the distribution of colors, artificially moved to $z=0.3$, of the SDSS galaxies in the spectroscopic catalog, and (ii) the distribution of apparent colors of the SDSS galaxies in the photometric catalog. In these comparisons, (i) shows what color distribution the galaxies at $z=0.3$ will have (that is, those galaxies are the members of an artificially-built $z=0.3$ cluster), while (ii) shows what the color distribution of the non-cluster field galaxies looks like (that is, they are contamination). To minimize the field-to-field variation effect for field galaxy properties, we use the SDSS galaxies in the $3^\circ\times3^\circ$ field (except the central $10'\times10'$ area; from the SDSS photometry catalog) centered on {\mc} without K-correction, for (ii).

Figure~\ref{comprel} shows an example for the estimation of completeness and reliability in our CMR selection. The magnitude range in Figure~\ref{comprel} ($20.0<i<22.0$ or $-20.4<M_i<-18.4$ at $z=0.3$) is the faintest range where the SDSS photometric catalog is complete \citep[$5\sigma$ completeness at $i\lesssim 22$;][]{yor00}. The galaxies in our observation is the combination of the $z=0.3$ galaxies (cluster members) and field galaxies:
\begin{equation}
\mathbb{C}_{\textrm{\scriptsize{obs}}}
= f_{\textrm{\scriptsize{c}}}\; \mathbb{C}_{\textrm{\scriptsize{c}}}
+ f_{\textrm{\scriptsize{f}}}\; \mathbb{C}_{\textrm{\scriptsize{f}}},
\end{equation}
where $\mathbb{C}_{\textrm{\scriptsize{obs}}}$, $\mathbb{C}_{\textrm{\scriptsize{c}}}$, and $\mathbb{C}_{\textrm{\scriptsize{f}}}$ are the color distributions of all observed galaxies, cluster members, and field galaxies, respectively; and $f_{\textrm{\scriptsize{c}}}$ and $f_{\textrm{\scriptsize{f}}}$ are the fractions of cluster members and field galaxies, respectively.
Here, it is assumed that $\mathbb{C}_{\textrm{\scriptsize{c}}}$ is like (i) or the solid-lined distribution in Figure~\ref{comprel}, while $\mathbb{C}_{\textrm{\scriptsize{f}}}$ is like (ii) or the dashed-lined distribution in Figure~\ref{comprel}. 

Now, to estimate the completeness and reliability, it should be determined how many galaxies in $\mathbb{C}_{\textrm{\scriptsize{c}}}$ and $\mathbb{C}_{\textrm{\scriptsize{f}}}$ satisfy our CMR selection criteria and how many do not.
At $-20.4<M_i<-18.4$ (Figure~\ref{comprel}), the number of selected cluster members ($N_{in}$) is 38, while the number of the rejected ($N_{out}$) is 21. In addition, the number fraction of the galaxies satisfying our criteria at this magnitude range among $\mathbb{C}_{\textrm{\scriptsize{c}}}$ galaxies is $95\%$ by our selection of CMR criteria (see Section~\ref{memsel}), while that among $\mathbb{C}_{\textrm{\scriptsize{f}}}$ is measured to be $43\%$ in Figure~\ref{comprel}.

Now, $N_{in}$ and $N_{out}$ are expressed as follows:
{\setlength\arraycolsep{0pt}\begin{eqnarray}
& & N_{in}:N_{out} = 38:21 \nonumber \\
& & = 0.95 N_c + 0.43 N_f : 0.05 N_c + 0.57 N_f,
\end{eqnarray}}
where $N_c$ is the number of genuine cluster members and $N_f$ is the number of field (foreground or background) objects.
Then, $N_c$ and $N_f$ are simply calculated to be 24.29 and 34.71, respectively, and the reliability ($R$) is:
\begin{equation}
R = \frac{0.95 N_c}{0.95 N_c + 0.43 N_f}\approx0.6072.
\end{equation}
Thus, we approximate the completeness and reliability of our CMR selection at $-20.4<M_i<-18.4$ to be $95\%$ and $61\%$, respectively.
That is, at $-20.4<M_i<-18.4$, it is expected that we secure $95\%$ of all genuine cluster members with our CMR selection, but $39\%$ of the selected objects may not be genuine cluster members.

These values are only for the CMR selection. Among the CMR-selected 38 objects at $-20.4<M_i<-18.4$, 36 objects also satisfy the size and concentration criteria, while 2 objects do not. As we apply more selection criteria, the completeness tends to decrease, while the reliability is expected to increase, because more selection means possible rejection of more genuine members but rejection of even more non-members at the same time.
However, since there is no appropriate control sample unlike the CMR selection, it is difficult to accurately estimate the completeness and reliability after the additional selections for size and concentration. Thus, we simply estimate them under two extreme conditions.
In the ideal case (that is, the size and concentration criteria rejected only non-members perfectly), the completeness would not vary ($95\%$), whereas the reliability would increase from $61\%$ to:
\begin{equation}
R = \frac{0.95 N_c}{0.95 N_c + 0.43 N_f - 2} \approx 0.6410,
\end{equation}
that is, $64\%$.
On the other hand, if additional selections did not work at all and members and non-members are rejected randomly (that is, with $N_c:N_f\approx 24:35$ ratio), the reliability would not vary ($61\%$), while the completeness ($C$) would be reduced from $95\%$ to:
\begin{equation}
C = \frac{0.95 N_c - 2 \times \frac{24}{24+35}}{N_c} \approx 0.9165,
\end{equation}
that is, $92\%$.
In conclusion, the completeness and reliability for our final member selection at $-20.4<M_i<-18.4$ are expected to be $92-95\%$ and $61-64\%$, respectively.

\begin{figure}[t]
\plotone{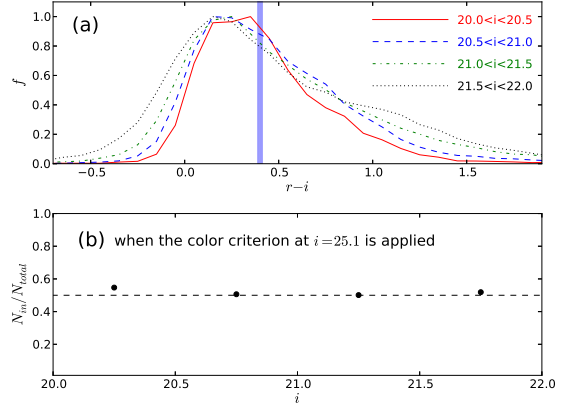}
\caption{ Approximating the color distribution of field galaxies at $-15.5<M_i<-15.0$ ($24.9<i<25.4$).
(a) The color distributions of field galaxies between $20.0<i<22.0$, divided into 4 magnitude bins with 0.5 magnitude interval. The shaded vertical stripe is the member selection criterion using the CMR at $24.9<i<25.4$, which is the faint end in our final sample. The shown criterion is the upper cut only, whereas the lower cut is not displayed in the plot range.
(b) The number fraction of galaxies satisfying the CMR criteria of $i=25.1$ ($M_i\approx-15.3$) along magnitude from $i=20$ to $i=22$. The horizontal dashed line indicates $N_{in}/N_{total}=0.5$.
\label{fieldcol}}
\end{figure}

For fainter galaxies, the same process is tried to estimate the completeness and reliability. However, the problem is that we can not secure $\mathbb{C}_f$ at $i>22$ due to the completeness limit of the SDSS photometric catalog.
Thus, we can only roughly approximate them for faint galaxies.
In Figure~\ref{fieldcol}(a), we found that the apparent color distribution of field galaxies changes mildly from $i\sim20$ to $i\sim22$, although it starts to be distorted at $i>22$ due to the sample incompleteness. Moreover, the number fraction of galaxies satisfying the CMR criterion of $i=25.1$ ($M_i\approx-15.3$) does hardly vary from $i=20$ to $i=22$, keeping $N_{in}/N_{total}\approx0.5$, as shown in Figure~\ref{fieldcol}(b).
Thus, we simply assume that the $N_{in}/N_{total}$ value does not significantly vary up to $i\sim25.4$ (corresponding to $M_i\sim-15$; the faintest limit in our sample). This assumption may not be strictly true, but we adopt it since we do not have any better alternative.

Under this assumption, the remaining processes are the same as those for bright galaxies.
At $-15.5<M_i<-15.0$, $N_{in}$ and $N_{out}$ are 64 and 38, respectively. The number fraction of the galaxies satisfying our criteria at this magnitude range among $\mathbb{C}_{\textrm{\scriptsize{c}}}$ galaxies is $95\%$ by our selection of CMR criteria, while that among $\mathbb{C}_{\textrm{\scriptsize{f}}}$ is assumed to be $50\%$. These conditions result in $N_c=28.89$ and $N_f=73.11$, and thus $C=95\%$ and $R=43\%$.
Among the CMR-selected 64 galaxies, 31 galaxies are rejected in the size and concentration criteria. As done for bright galaxies, we simply estimate $C$ and $R$ under two extreme conditions. In the ideal case (size and concentration selections work perfectly), $C$ does not vary, while $R$ increases to $83\%$. In the bad case (size and concentration selections work randomly), $R$ does not change, but $C$ decreases to $65\%$.
In summary, at $-15.5<M_i<-15.0$, $C$ is expected to be $65 - 95\%$, whereas $R$ is expected to be $43 - 83\%$.

Note that the upper limit in reliability at $-15.5<M_i<-15.0$ is even larger than that at $-20.4<M_i<-18.4$. This does not imply that the reliability at the faint end can be better than that at the bright end. Those values are just \emph{arithmetic} upper-limits\footnote{To achieve the upper limit, the additional selection using the SMR and ConMR should work \emph{perfectly}, which is not plausible.}, which simply reflects the large uncertainty in the reliability estimation. That is, the reason that the upper limit in reliability at the faint end looks high is just because the uncertainty is large. We speculate that the actual reliability (and similarly the actual completeness) at $-15.5<M_i<-15.0$ may be lower than that at $-20.4<M_i<-18.4$, or at most similar.

Additionally, we cross-checked our CMR selection, using $\sim$ 10,000 SDSS galaxies with photo-z's randomly selected at $20<i<21$, where the uncertainty in photo-z estimation is smaller than 0.1. We divided those SDSS galaxies into two groups: photo-z members ($0.2 \le$ photo-z $\le 0.4$) and photo-z non-members (photo-z $<0.2$ or photo-z $>0.4$). Note that this is a very rough division, but more accurate division is not effective due to the large uncertainty in photo-z estimation. Among the 3661 photo-z members, 1694 objects ($46\%$) satisfy our CMR criterion, while among the 3696 objects satisfying our CMR criterion, 1694 objects ($46\%$) are photo-z members. That is, in this test, the $C$ and $R$ of our CMR selection at $20 < i < 21$ are estimated to be both $46\%$.
These values are much smaller than our original estimates, but it does not necessarily mean that the original values of $C$ and $R$ are significantly overestimated. Rather, it is understood as the result from the difference in color distribution between photo-z galaxies (at photo-z $=0.3\pm0.1$) and spectroscopic redshift (spec-z) galaxies (exactly at $z = 0.3$). Although the colors at $z = 0.3$ of spec-z galaxies were re-calculated through k-correction, the color distribution of spec-z galaxies is expected to better reflect the actual color distribution of galaxies in a cluster at $z=0.3$ than that of photo-z galaxies does.

There are several caveats in our completeness and reliability estimation.
First, the SDSS spectroscopic catalog includes both cluster and field galaxies, not purely cluster galaxies, which may result in inaccurate estimation of $C$ and $R$. Since the color distribution of galaxies in a galaxy cluster is typically narrower than that in low-density fields, the actual membership reliability may be lower. Nevertheless, since it is difficult to determine a \emph{typical} color distribution in a galaxy cluster due to the cluster-to-cluster variation, we used such a mixed distribution from the SDSS.
Second, similarly, our field galaxy control sample from the SDSS photometric catalog may not be perfect. Although we use the surrounding $3^{\circ}\times3^{\circ}$ field to minimize the field-to-field variation, there still exists a possibility that the actual color distribution of background and foreground galaxies in our observed field is different from that of the control field galaxies. Moreover, even some galaxies in the control field may have redshifts of $\sim0.3$, which is an uncertainty factor in our $C$ and $R$ estimation. However, the number of galaxies in our control field is as many as 136,000 at $i\le22$, while all detected galaxies at $i\le22$ in our analysis area are only 90 in number (before member selection). This implies that even if there are several more galaxy clusters with $z=0.3$ in our control field, their effect will be very limited.
Finally, the assumption that the field galaxy color distribution does not vary significantly at $i>22$ is just an assumption, which is not guaranteed. This makes the uncertainty large in the estimation of $C$ and $R$ for faint galaxies.

\subsection{Local Environmental Parameters}\label{envpar}

\begin{figure}[t]
\plotone{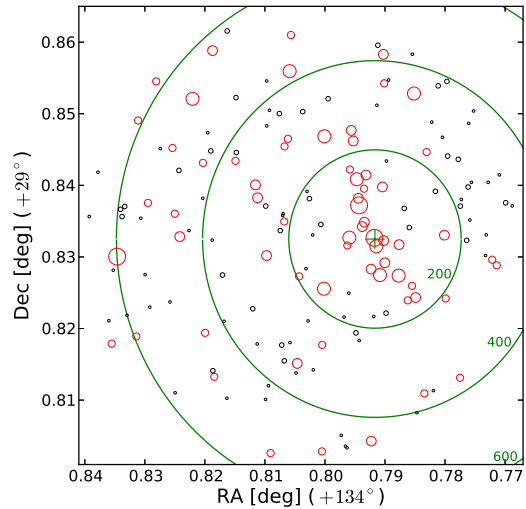}
\caption{ Spatial distribution of the selected cluster members. The size of each symbol is proportional to the luminosity of each object. Red open circles are bright ($M_i\le-18$) members, while black open circles are faint ($-18<M_i\le-15$) members. The green cross denotes the center of {\mc}, and the green concentric circles show the cluster-centric radii of $45''$, $90''$ and $135''$, which respectively correspond to 200 kpc, 400 kpc and 600 kpc at $z=0.3$.
\label{spatial}}
\end{figure}

\begin{figure}[t]
\plotone{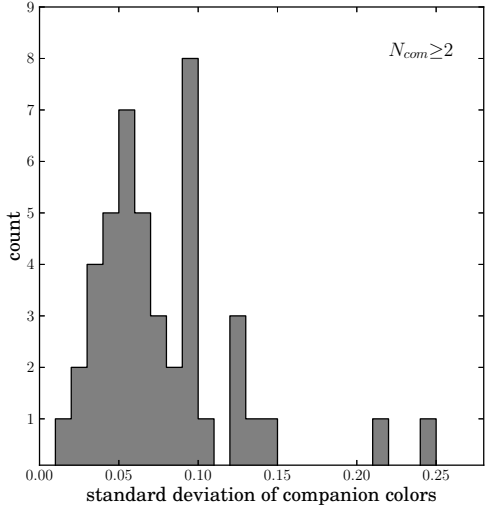}
\caption{ The distribution of the standard deviation in faint companion colors for a given bright galaxy. Since the standard deviation for a single companion galaxy ($N_{\textrm{\scriptsize com}}=1$) is zero, the histogram is shown only for 45 bright galaxies with $N_{\textrm{\scriptsize com}}\ge2$. The standard deviation in color of all 91 faint galaxies in our sample is as large as 0.79.
\label{satdev}}
\end{figure}

Figure~\ref{spatial} displays the spatial distribution of the selected cluster member galaxies. To quantitatively estimate the relationship between bright galaxies and their faint companions, we need to parameterize the physical properties of faint galaxies around each bright galaxy. This process is almost the same as that to estimate the local environmental parameters.
The basic idea is to calculate the mean or integrated quantities of galaxies around each bright galaxy, where those quantities may be number, luminosity, color, and so on. In this process, we adopt a spline kernel with $22.4''$ ($\sim 100$ kpc at $z=0.3$) radius for smoothing. For example, the luminosity density ($\Sigma$) is defined as:
\begin{equation}
\Sigma = \sum_{k=1}^{n}
f_{sp}(d_k) \times 10^{-0.4 M_k}/A,
\end{equation}
where $n$, $f_{sp}$, $d_k$, $M_k$, and $A$ are the number of galaxies in the local area, spline kernel function, projected distance to the local area center, absolute magnitude, and local area, respectively.
For more details about the estimation of environmental parameters, see also Appendix A of \citet{lee10}.
In this paper, $i$-band local luminosity density is used as a parameter representing the local environment, instead of local number density that does not reflect various luminosities ($\sim$ various masses) of galaxies.

Since the spatial resolutions of our images are poor (up to $2''$ or 9 kpc), the only practical quantity to be parameterized for faint galaxies is color. We estimated the weighted mean $r-i$ color of faint companions ($(r-i)_{\textrm{\scriptsize com}}$) around each bright galaxy, as follows:
{\setlength\arraycolsep{0pt}\begin{eqnarray}
& & (r-i)_{\textrm{\scriptsize com}} = \nonumber \\
& & \frac{\sum_{k=1}^{n} f_{sp}(d_k) \times f_{L}(k) \times (r-i)_k}{\sum_{k=1}^{n} f_{sp}(d_k) \times f_{L}(k)},
\end{eqnarray}
where $(r-i)_k$ is the color of each faint galaxy within $22.4''$ from a given bright cluster galaxy. The \emph{faint} galaxies mean the selected cluster members with $-18< M_i \le-15$. In addition to the spline kernel weight according to the distance to the adjacent bright galaxy, the luminosity weight ($f_{L}$) is also applied, which is defined as:
\begin{equation}
f_{L}(k) = 10^{-0.4 \big(M_i(k)+18\big)},
\end{equation}
where $M_i(k)$ is the $i$-band absolute magnitude of a given faint galaxy.
Since fainter galaxies have larger photometric uncertainties, the luminosity weight is useful to reduce the influence of faint objects (with large uncertainties) on the weighted mean color of companions.

Figure~\ref{satdev} shows how dispersed the companion colors for a given bright galaxy. For bright galaxies with the number of faint companions larger than 1 ($N_{\textrm{\scriptsize com}}\ge2$), the standard deviation of companion colors are mostly smaller than 0.15 (only 2 exceptions). Moreover, about $80\%$ of $N_{\textrm{\scriptsize com}}\ge2$ bright galaxies have companion color deviations smaller than 0.1, which is approximately as small as the photometric errors of faint ($-18<M_i\le-15$) galaxies.
Since the standard deviation in color of all 91 faint galaxies in our sample is as large as 0.79, the result in Figure~\ref{satdev} implies that faint companions around a given bright galaxy tend to have quite consistent photometric properties.

\begin{deluxetable}{lr @{.} lr @{.} lr @{.} l}
\tablenum{1} \tablecolumns{7} \tablecaption{ Statistics of Faint Companions for Each Bright Galaxy\label{satstat}} \tablewidth{0pt}
\tablehead{ & \multicolumn{2}{c}{$N_{\textrm{\scriptsize com}}\ge0$} & \multicolumn{2}{c}{$N_{\textrm{\scriptsize com}}\ge1$} & \multicolumn{2}{c}{$N_{\textrm{\scriptsize com}}\ge2$} }
\startdata
The number of bright galaxies & \multicolumn{2}{l}{65} & \multicolumn{2}{l}{60} & \multicolumn{2}{l}{45} \\
Mean $N_{\textrm{\scriptsize com}}$ $^{(i)}$ & $\;\:$2 & 66 & 2 & 88 & 3 & 51 \\
Mean Separation (in arcsec) $^{(ii)}$ & \multicolumn{2}{c}{---} & 16 & 2 & 15 & 6 \\
Mean Separation (in $R_{50,\textrm{\scriptsize bri}}$) & \multicolumn{2}{c}{---} & 12 & 8 & 12 & 9 \\
Minimum Separation (in arcsec) & \multicolumn{2}{c}{---} & 12 & 8 & 11 & 1 \\
Minimum Separation (in $R_{50,\textrm{\scriptsize bri}}$) & \multicolumn{2}{c}{---} & 9 & 9 & 9 & 0 \\
\hline \hline
($+$ photo-z) & \multicolumn{2}{c}{$N_{\textrm{\scriptsize com}}\ge0$} & \multicolumn{2}{c}{$N_{\textrm{\scriptsize com}}\ge1$} & \multicolumn{2}{c}{$N_{\textrm{\scriptsize com}}\ge2$} \\
\hline
The number of bright galaxies & \multicolumn{2}{l}{70} & \multicolumn{2}{l}{65} & \multicolumn{2}{l}{50} \\
Mean $N_{\textrm{\scriptsize com}}$ & $\;\:$2 & 69 & 2 & 89 & 3 & 46 \\
Mean Separation (in arcsec) & \multicolumn{2}{c}{---} & 16 & 0 & 15 & 5 \\
Mean Separation (in $R_{50,\textrm{\scriptsize bri}}$) & \multicolumn{2}{c}{---} & 12 & 9 & 13 & 0 \\
Minimum Separation (in arcsec) & \multicolumn{2}{c}{---} & 12 & 5 & 10 & 9 \\
Minimum Separation (in $R_{50,\textrm{\scriptsize bri}}$) & \multicolumn{2}{c}{---} & 9 & 8 & 9 & 0 \\
\hline \hline
($-$ photo-z) & \multicolumn{2}{c}{$N_{\textrm{\scriptsize com}}\ge0$} & \multicolumn{2}{c}{$N_{\textrm{\scriptsize com}}\ge1$} & \multicolumn{2}{c}{$N_{\textrm{\scriptsize com}}\ge2$} \\
\hline
The number of bright galaxies & \multicolumn{2}{l}{57} & \multicolumn{2}{l}{52} & \multicolumn{2}{l}{37} \\
Mean $N_{\textrm{\scriptsize com}}$ & $\;\:$2 & 49 & 2 & 73 & 3 & 43 \\
Mean Separation (in arcsec) & \multicolumn{2}{c}{---} & 16 & 3 & 15 & 7 \\
Mean Separation (in $R_{50,\textrm{\scriptsize bri}}$) & \multicolumn{2}{c}{---} & 12 & 7 & 12 & 8 \\
Minimum Separation (in arcsec) & \multicolumn{2}{c}{---} & 13 & 1 & 11 & 2 \\
Minimum Separation (in $R_{50,\textrm{\scriptsize bri}}$) & \multicolumn{2}{c}{---} & 10 & 0 & 8 & 9 \\
\enddata
\tablecomments{$(i)$ $N_{\textrm{\scriptsize com}}$ is the number of faint companions for a given bright galaxy.
$(ii)$ \emph{Separation} means the distance between a bright galaxy and its faint companion.}
\end{deluxetable}

Several collective properties of faint companions for each bright galaxy are listed in Table~\ref{satstat}.
Among the 65 bright galaxies, 5 galaxies have no companion with $-18<M_i\le-15$, which are excluded in the subsequent analysis. The number of faint companions for each bright galaxy is not large: 2.66 on average for all 65 bright galaxies or 2.88 on average for 60 bright galaxies with at least one faint companion. The mean separation between bright galaxies and their faint companions is $15-16''$ (corresponding to $67-71$ kpc at $z=0.3$) or $12-13\,R_{50,\textrm{\scriptsize bri}}$, where $R_{50,\textrm{\scriptsize bri}}$ is the half-light radius of each bright galaxy.

Note that each faint companion may be adjacent to multiple bright galaxies in our definition, which is the main reason that the weight for separation is necessary in estimating $(r-i)_{\textrm{\scriptsize com}}$. When one faint galaxy is located within 100 kpc from two or more bright galaxies, it is difficult to distinguish which one genuinely hosts that faint galaxy, not only because we are just looking at the projected distribution of galaxies but also because we do not know the dynamics of those galaxies. However, the probability that a faint galaxy is a genuine satellite galaxy of a bright galaxy decreases with increasing distance between them. Based on this idea, even though a faint galaxy is adjacent to multiple bright galaxies, its contribution to the $(r-i)_{\textrm{\scriptsize com}}$ value of each bright galaxy varies depending on its distance from each one, which is expected to partially reflect the probability of a real host-satellite relationship.
However, for bright galaxies with a single companion, the \emph{weight} does not work actually, and thus the risk of contamination is higher than that for bright galaxies with multiple companions. Hence, in the subsequent analysis, we investigate both the cases of $N_{\textrm{\scriptsize com}}\ge1$ and $N_{\textrm{\scriptsize com}}\ge2$.

\section{RESULTS}

\begin{figure*}
\includegraphics[width=1.0\textwidth]{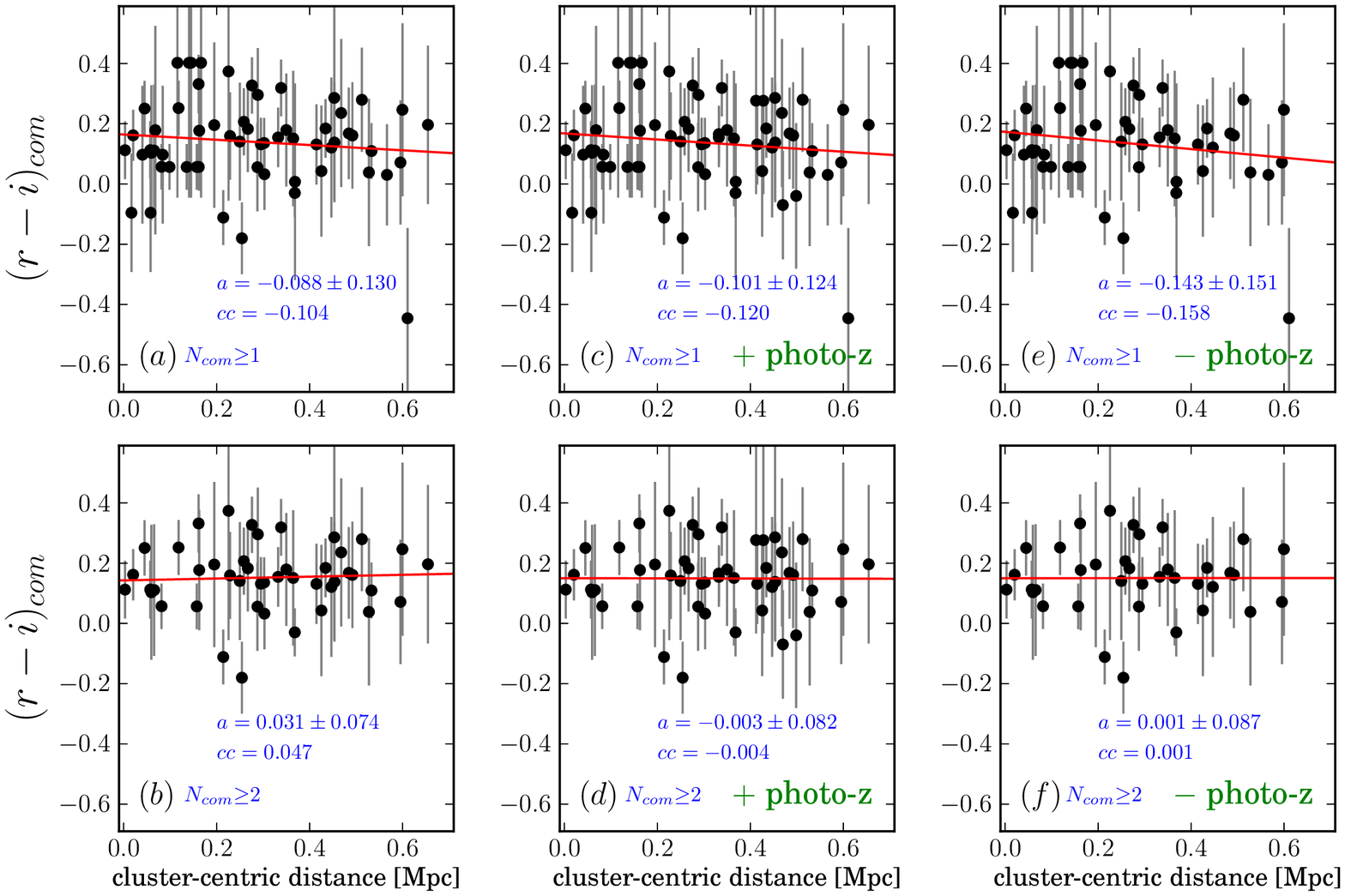}
\caption{ The correlation between the mean colors of faint companions and the cluster-centric distances of their adjacent bright galaxies. Upper and lower panels display the results for different limits in the number of faint companions for a given bright galaxy: (a) $N_{\textrm{\scriptsize com}}\ge1$, and (b) $N_{\textrm{\scriptsize com}}\ge2$. Panels (c) and (d) are the same as (a) and (b) respectively, except that the photo-z members are added to the main sample ($+$ photo-z), while (e) and (f) are for the sample in which the photo-z non-members are removed from the main sample ($-$ photo-z).
In each panel, the slope of linear least-squared fit ($a$ value), the Bootstrap uncertainty for $a$ ($\pm$ value), and the correlation coefficient ($cc$ value) are denoted. Throughout Figures~\ref{cldist} to \ref{hostcol}, the number of points is 60, 45, 65, 50, 52, and 37 in panels (a), (b), (c), (d), (e), and (f), respectively. No significant dependence on cluster-centric distance is found.
\label{cldist}}
\end{figure*}

\begin{figure*}
\includegraphics[width=1.0\textwidth]{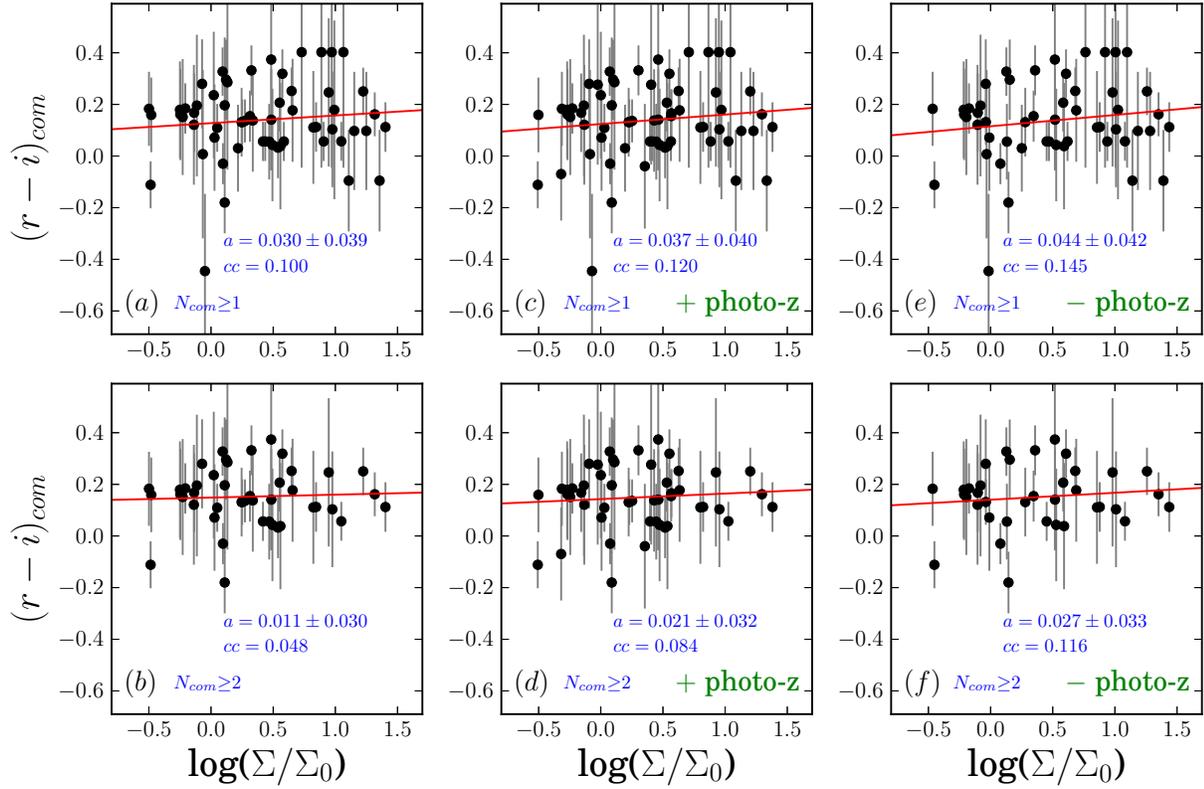}
\caption{ The correlation between the mean colors of faint companions and the $i$-band luminosity density ($\Sigma$) around their adjacent bright galaxies. The luminosity density was normalized by the mean luminosity density of the entire analysis area ($\Sigma_0$). No significant dependence on local luminosity density is found.
\label{lden}}
\end{figure*}

\begin{figure*}
\includegraphics[width=1.0\textwidth]{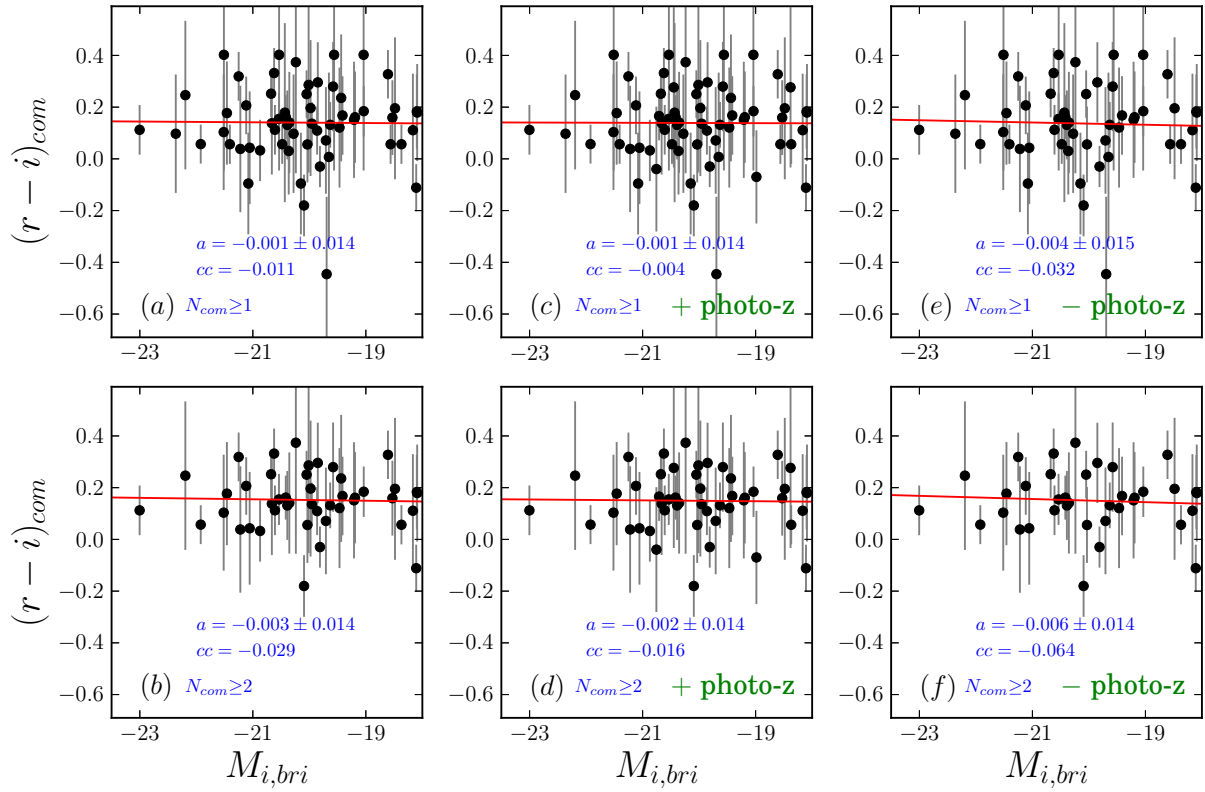}
\caption{ The correlation between the mean colors of faint companions and the $i$-band absolute magnitudes of their adjacent bright galaxies. No significant correlation is found.
\label{hostmag}}
\end{figure*}

\begin{figure*}
\includegraphics[width=1.0\textwidth]{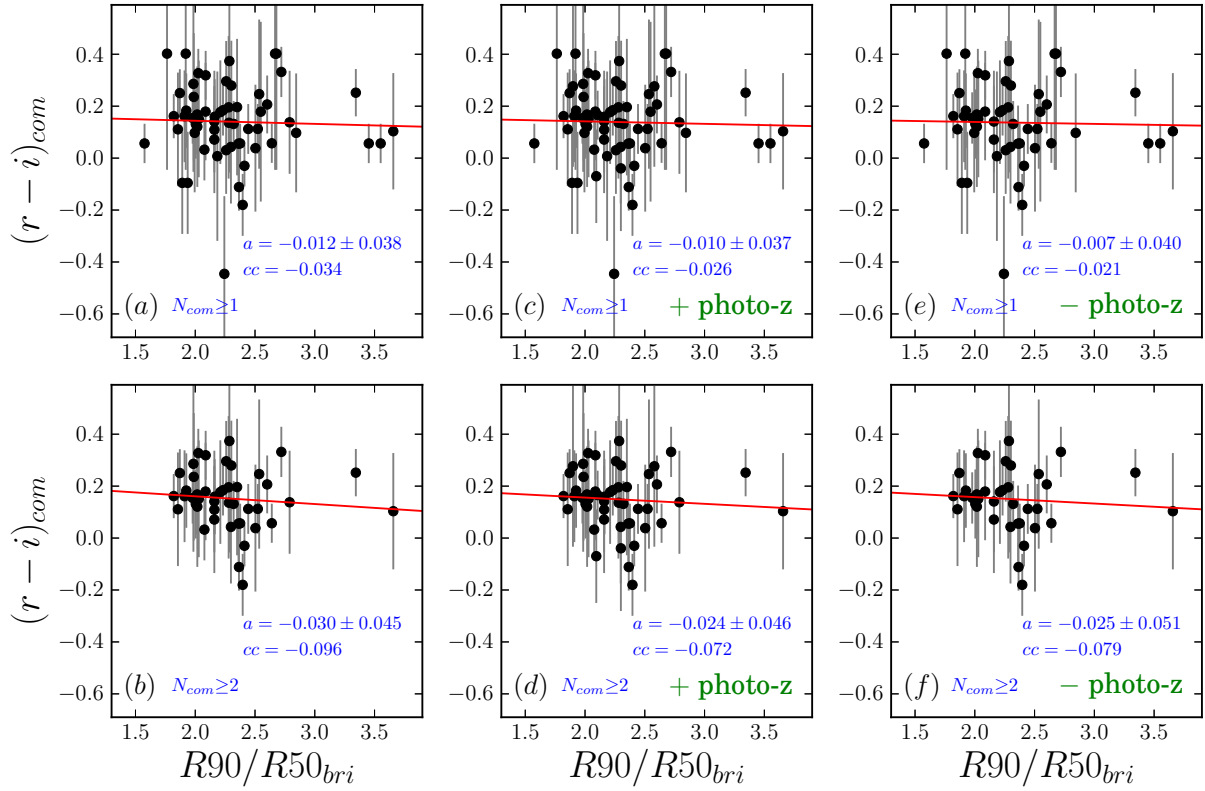}
\caption{ The correlation between the mean colors of faint companions and the $i$-band light concentrations (R90/R50) of their adjacent bright galaxies. No significant dependence on local luminosity density is found.
\label{hostcon}}
\end{figure*}

\begin{figure*}
\includegraphics[width=1.0\textwidth]{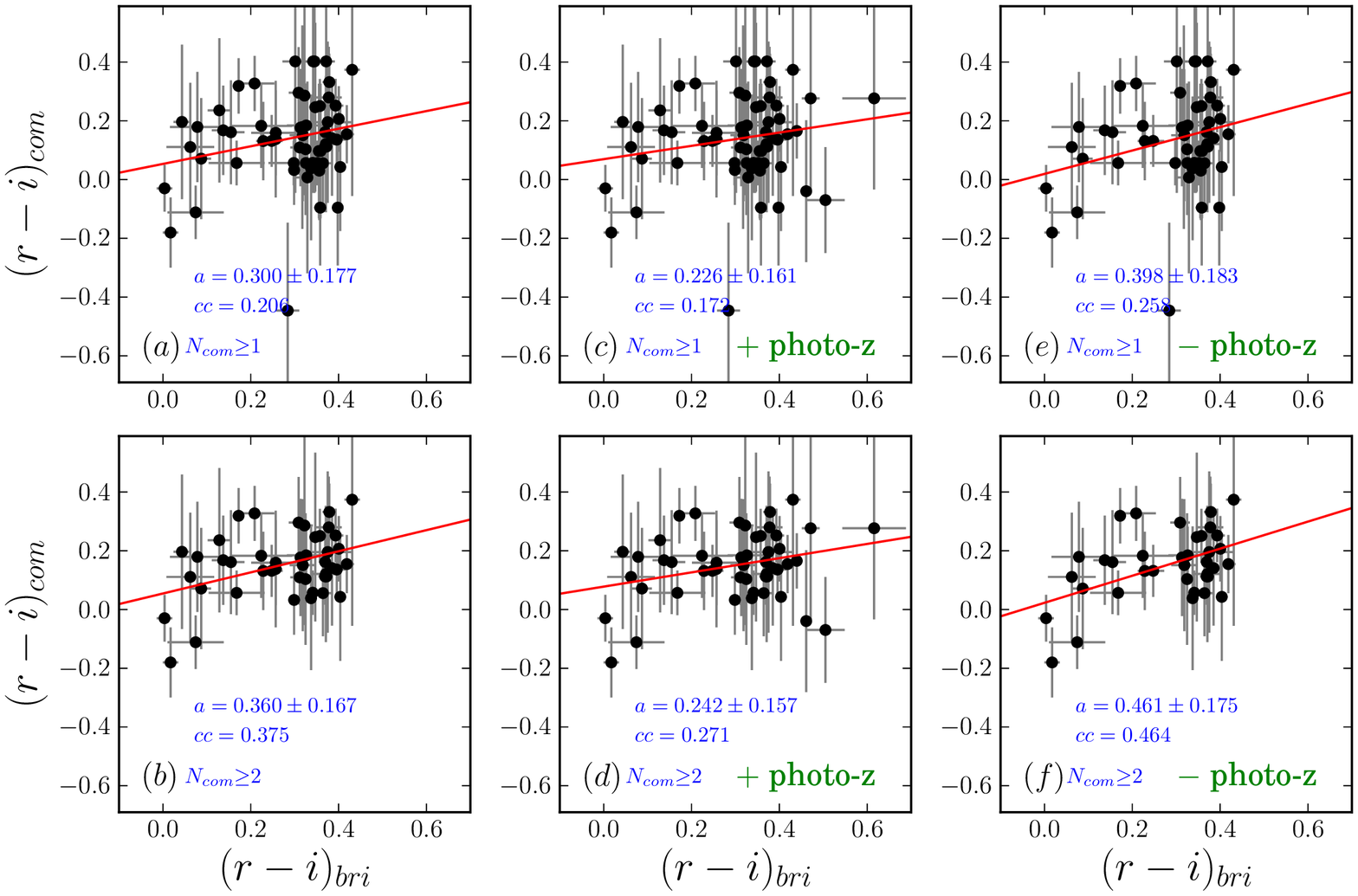}
\caption{ The correlation between the mean colors of faint companions and the colors of their adjacent bright galaxies. Unlike the previous plots, the correlation slopes are marginally detected in panels (b), (e), and (f) ($2.2\sigma - 2.6\sigma$ to the Bootstrap uncertainties).
\label{hostcol}}
\end{figure*}

The main goal of this paper is to investigate the relationship between bright galaxies and their faint companions in a galaxy cluster. To achieve this goal, we need to first check the environmental effects, because the cluster environment is known to strongly affect the properties of galaxies in a large scale \citep[e.g.,][]{par07,ada12,pim12}. From numerous previous studies, now it is well known that galaxies in high-density environments or galaxy clusters tend to be redder than galaxies in low density environments or fields \citep{bam09,han09,wei10,cib13}. However, such trends are not obvious in our sample, as shown in Figures~\ref{cldist} and \ref{lden}.

Figure~\ref{cldist} presents the dependence of $(r-i)_{\textrm{\scriptsize com}}$ on cluster-centric distance, in which no meaningful trend is found. The estimated correlation slopes are very small and not significant ($<1\sigma$) to the Bootstrap uncertainties.
This may be partially due to the color integration within 100 kpc radius, which can wash out the color dependence on local environment. However, even when the dependence on cluster-centric distance for the individual colors of faint galaxies is tested, its significance to the Bootstrap uncertainty is just $\sim 1.5\sigma$, which is still statistically insignificant.
Similarly, $(r-i)_{\textrm{\scriptsize com}}$ does not show any significant dependence on local luminosity density in Figure~\ref{lden}. These results do not necessarily indicate that the colors of faint companions are not affected by local environments, because the uncertainty is large due to possible contamination. However, at least in our sample, the local environmental effect is not detected.

Since the local environments of the cluster do not significantly affect the colors of faint galaxies, now we test the effect of adjacent bright galaxies on $(r-i)_{\textrm{\scriptsize com}}$.
Figure~\ref{hostmag} shows the relationship between $(r-i)_{\textrm{\scriptsize com}}$ and the bright galaxy luminosities, in which no statistically significant dependence is found.
As presented in Figure~\ref{hostcon}, the dependence on the light concentrations of bright galaxies is not significant, either. These results show the luminosities and light profiles of bright galaxies do not significantly influence the colors of their faint companions, at least in our sample.

Finally, the dependence of $(r-i)_{\textrm{\scriptsize com}}$ on the colors of bright galaxies is tested in Figure~\ref{hostcol}. Unlike the 4 parameters in the previous plots, the slopes are measurable, in the sense that blue bright galaxies tend to have blue faint companions. Particularly for the sample with $N_{\textrm{\scriptsize com}}\ge2$ (Figure~\ref{hostcol}(b)), the slope ($\Delta(r-i)_{\textrm{\scriptsize com}}/\Delta(r-i)_{\textrm{\scriptsize bri}}=0.360$) has $2.2\sigma$ significance to the Bootstrap uncertainty, and for the `$-$ photo-z' sample with $N_{\textrm{\scriptsize com}}\ge2$ (Figure~\ref{hostcol}(f)), the significance is even as large as $2.6\sigma$. These values are statistically not significant ($<3\sigma$) but marginal ($>2\sigma$). The correlation coefficients are as large as 0.375 and 0.464 in Figure~\ref{hostcol}(b) and \ref{hostcol}(f), respectively.
Whereas the statistical significances are mostly smaller than $1\sigma$ for the other quantities (cluster-centric distance, local luminosity density, bright galaxy luminosity, and bright galaxy light concentration), the bright galaxy color appears to considerably influence the colors of faint companions in our results.

\section{DISCUSSION}

\subsection{Possible Uncertainties}\label{uncer}

Basically, it is very difficult to determine membership without velocity information in studies of galaxy clusters. Thus, any method only based on image data may have large uncertainty in member selection, which affect the results. Here, several factors that may mislead the results in our analysis are listed and their actual effects are discussed.

The first is the low reliability in our member selection. According to our estimation in Section~\ref{comrel}, the reliability is minimum $43\%$ and maximum $83\%$ at $-15.5<M_i<-15.0$, which means that almost 3 of 5 cluster members are falsely selected in the worst case. However, as mentioned in Section~\ref{memsel}, falsely-selected members are expected to make the trends or correlations in our results more unclear by mixing out the signal with scattered noise, because falsely-selected members must be foreground or background objects with random properties. It is not reasonable that their properties have any correlation with their close (in projected distance) bright galaxies. That is, the low reliability in member selection does not weaken the significance of our finding; on the contrary, our results might become even more obvious if we selected cluster members better.

Second, the deblending process of the \emph{Source Extractor} may not have worked perfectly. That is, small fluctuations in the surface profile of a bright galaxy may have been regarded as faint companion galaxies, which can cause a false correlation between the colors of bright galaxies and their `faint companions'. If the `faint companions' were over-deblended ambient light of a bright galaxy, it would be natural that they had colors similar to that of the bright galaxy. Although our member selection method is expected to have removed a significant number of those over-detected light fluctuations, it may not be perfect.
However, as shown in Figure~\ref{spatial}, most faint companions are not so close to the bright galaxies. As mentioned in Section~\ref{envpar}, the mean distance from faint companions to their adjacent bright galaxies is as large as $15''$, which is typically 12 times larger than the half-light radii of the bright galaxies. Because a faint companion closer to its adjacent bright galaxy has a larger weight in the $(r-i)_{\textrm{\scriptsize com}}$ calculation, we additionally checked the distance to the closest faint companion from each bright galaxy, finding that its mean value is $7.6''$ or $6.4\,R_{50,\textrm{\scriptsize bri}}$, which is not a very small distance, either.
Hence, even if falsely-deblended objects are included in our final sample, it may be very rare, not enough to critically influence our results.

Third, if redder bright galaxies have preferentially brighter companions around them, then the companion galaxies may naturally have redder colors due to the well-known color-magnitude relation. To check this possibility, we tested the relationship between the colors of bright galaxies and the mean luminosities of their faint companions.
The mean luminosity of faint companions ($L_{i,\textrm{\scriptsize com}}$) is defined as:
\begin{equation}
L_{i,\textrm{\scriptsize com}} = \frac{\sum_{k=1}^{n} f_{sp}(d_k) \times L_i(k)}{\sum_{k=1}^{n} f_{sp}(d_k)},
\end{equation}
where $L_i(k)$ is the luminosity of each faint companion galaxy. We converted $L_{i,\textrm{\scriptsize com}}$ into absolute magnitude ($M_{i,\textrm{\scriptsize com}}$) and estimated its dependence on the bright galaxy color in a way similar to those of Figures~\ref{cldist} - \ref{hostcol}.
As a result, we found that the correlation slope bewteen $(r-i)_{\textrm{\scriptsize bri}}$ and $M_{i,\textrm{\scriptsize com}}$ is $a = 0.292\pm0.177$ with a correlation coefficient of $cc = 0.039$ for the $N_{\textrm{\scriptsize com}} \ge 1$ sample. This result is not so different for the $N_{\textrm{\scriptsize com}} \ge 2$ sample: $a = 0.227\pm0.167$ with $cc = 0.038$. That is, we can not find any meaningful correlation between the bright galaxy color and the companion galaxy luminosity. Note that even if we ignore the bootstrap uncertainty, the slope indicates the opposite relationship: redder bright galaxies tend to have less luminous companions.

Finally, the marginal `signals' in Figure~\ref{hostcol} may be results of coincidence, due to small-number statistics. Since the total number of our selected cluster members at $M_i<-15$ is only 156, this concern may be reasonable. To resolve this issue, we carried out permutation tests \citep[e.g.,][]{goo94}, which are statistical tests suitable for distinguishing whether a correlation exists between two variables: in this case, the colors of bright galaxies and the weighted mean colors of their faint companions. The brief procedure of this test is as follows. First, we shuffled the colors of all faint galaxies in our analysis area and calculated the weighted mean color of (shuffled) faint companions for each bright galaxy. Subsequently, from this random sampling, Figures~\ref{cldist} -- \ref{hostcol} were redrawn and their correlation coefficients were measured. After repeating such a random sampling 1000 times, we estimated the confidence level for each correlation by comparing the original correlation 
coefficient with the distribution of correlation coefficients in the random samples. The results are summarized in Table~\ref{rand}. From this test, we confirm that the results in Figure~\ref{hostcol}(b) and \ref{hostcol}(f) are not by chance due to small-number statistics. The correlation coefficients in Figure~\ref{hostcol}(b) and \ref{hostcol}(f) are as large as 0.375 with a confidence level of $98.7\%$ and 0.464 with a confidence level of $99.3\%$, respectively, which obviously shows that the bright galaxies are correlated in color with their faint companions.

\subsection{Origin of the Color Correlation}

As shown in \citet{phi13}, bright host galaxies and their faint satellite galaxies show close relationships in isolated groups, which is not strange because a lot of evidence supports that host and satellite galaxies often influence each other \citep[e.g.,][]{lar11,cha13,pau13}. However, is it possible that such close relationships are sustained even in the harsh environment of a galaxy cluster, in which direct galaxy-galaxy interactions hardly affect galaxy properties \citep{mer84,par09}?
In our results, the answer seems to be a cautious ``yes''. That is, the close relationships in photometric properties between bright galaxies and their faint companions are marginally detected even in a galaxy cluster.
In Section~\ref{uncer}, we discussed about the possibilities that these results were not real, but several possible uncertainties in our analysis are not likely to artificially produce the correlations.

If we accept that the correlations are real, our results may be interpreted in three ways.
The first is that a galaxy group consisting of a massive galaxy and its low-mass satellites may survive for a long time even after the group falls into a galaxy cluster. Today, more and more evidence is being found, supporting the idea that massive galaxy clusters have grown by merging smaller groups and that galaxies have rapidly evolved in the galaxy group stage \citep[so-called `pre-processing';][]{zab98,bal00,hoy12,vij13}.
Small groups of galaxies fallen into a galaxy cluster are expected to be eventually broken up by the strong tidal force in the gravitational potential of the galaxy cluster after sufficient time passes, and thus this scenario predicts that the color correlation between bright galaxies and their faint companions will be stronger in dynamically younger clusters.
Hence, to confirm this scenario, galaxy clusters in various dynamical stages need to be compared.
In addition, the trend along cluster-centric distance should be checked, too. Since the galaxies (or galaxy groups) at cluster outskirt may be fallen into the cluster in relatively recent times, if this scenario is true, then the color correlations are expected to be stronger at cluster outskirt.

Another scenario is that a massive galaxy may capture low-mass galaxies coming close to it in a galaxy cluster, although massive neighbors just pass by \citep{bas98,din00,ber03}. That is, the current companions may not be {\it in-situ} satellites of the bright galaxies.
Those newly-captured low-mass satellites may interact with the massive galaxy, resulting in the color relationship between them.
However, this picture should be checked in two major aspects. One is how efficient such capturing of low-mass galaxies by a massive galaxy is in a galaxy cluster. Based on simple numerical simulations, \citet{bas98} showed that dwarf galaxies in a galaxy cluster are captured by massive galaxies up to $5\%$, but more elaborate and diverse tests are needed for the comparisons with the real galaxy clusters under various conditions.
The other is what process makes the close relationships in photometric properties between a massive galaxy and its low-mass (captured) satellites, which requires changes in photometric properties of those galaxies after a capturing event. Even though direct tidal interactions between galaxies hardly happen in a galaxy cluster, hydrodynamic interactions may make it possible, in which not stars but gas in a galaxy moves to a close neighbor galaxy and causes additional star formation there \citep{par09}.
Nevertheless, since a typical cluster center is known to be a hostile environment where most galaxies lose their gas and thus stop star formation activities \citep[e.g., due to ram-pressure stripping;][]{gun72}, the plausibility of this scenario should be tested further, particularly in the sense that how much gas remains in the cluster galaxies for hydrodynamic interactions.

The last scenario is that a considerable number of the faint companions may have been tidally torn out from their adjacent bright galaxies. In other words, those companions may have been originally outer parts of bright galaxies, but may be separated by tidal force of another massive galaxy or the galaxy cluster itself \citep[so-called `tidal dwarf galaxies'; e.g.,][]{duc98,bou06,she09,kav12}. This scenario explains well why the colors of bright galaxies and their faint companions show good correlations.
From a study using the SDSS data \citep{kav12}, it was reported that the median separation between a host galaxy and its tidal satellites is $\sim4.5\, R_{50,\textrm{\scriptsize{host}}}$ and $95\%$ of tidal satellites are within $15\, R_{50,\textrm{\scriptsize{host}}}$.
Their results show that the mean separation between tidal dwarfs and their host galaxies is smaller than that between the bright galaxies and their faint companions in our result ($\sim12\,R_{50,\textrm{\scriptsize{bri}}}$). Thus, our faint companions with large separations may not be tidal dwarfs, but a considerable number of our faint companions seem to be within the tidal dwarf domain \citep[the $95\%$ range given in][]{kav12}.
However, this scenario also needs to be tested by checking how frequently such \emph{tidal tearing} events happen in a galaxy cluster.

\section{CONCLUSION}

We carried out a deep two-band photometric study of {\mc}, a galaxy cluster at $z=0.30$, to investigate the relationship between bright ($M_i \le -18$) galaxies and their faint ($-18<M_i \le -15$) companions in this galaxy cluster. While the weighted mean color of faint companion galaxies hardly depends on local environmental parameters (cluster-centric distance and local luminosity density) as well as the luminosity and concentration of bright galaxies ($<1\sigma$ significance) at least in our sample, it shows marginal dependence on the color of bright galaxies ($\sim2.2\sigma$ significance) for the $N_{\textrm{\scriptsize com}}\ge2$ sample. The statistical significance increases to $\sim2.6\sigma$ if we additionally remove non-members using the SDSS photometric redshift information from the main sample.
After several possible uncertainties were discussed, we concluded that it is not plausible for those uncertainties to coincidentally produce the results in this paper.
Using permutation tests, we confirmed that the correlation in color between bright galaxies and their faint companion is statistically reliable with a confidence level of $98.7\%$, when the sample is limited to $N_{\textrm{\scriptsize com}}\ge2$, or with $99.3\%$ when photo-z non-members are removed.

We suggest three scenarios to interpret our results:
\begin{enumerate}
 \item A galaxy group consisting of a massive galaxy and its low-mass satellites may survive for a long time even after the group falls into a galaxy cluster. Thus, the close relationship between bright galaxies and their faint companions would be the \emph{vestige of infallen groups}. To confirm this, it should be shown how the color correlations depend on the dynamical stages of clusters and the cluster-centric distances.
 \item A massive galaxy may capture low-mass galaxies coming close to it in a galaxy cluster, whereas massive neighbors just pass by. To confirm this scenario, it should be checked how efficient such \emph{capturing} is and how significantly the hydrodynamic interaction after capturing affects the colors of bright galaxies and their companions
 \item A considerable number of the faint companions may have been tidally torn out from bright galaxies. The efficiency of such \emph{tearing-out} should be tested.
\end{enumerate}

To confirm any of the suggested scenarios, similar investigation for a larger sample of galaxy clusters with deep imaging is required as well as numerical simulations with resolution high enough to distinguish faint dwarf galaxies. If the first scenario is true and the color correlation between bright galaxies and their faint companions becomes weaker as time goes after group infalling (possibly by loss of satellites due to cluster tidal force), then it is expected that the color correlation will disappear as a galaxy cluster is getting dynamically older. On the other hand, the color correlation may not disappear even after long time has passed in the second or third scenario, although currently we can not give detailed predictions for those.
To study further about this issue, we are analyzing more galaxy clusters using deeper images.

\acknowledgments

We appreciate the anonymous referee who motivated significant improvement of this paper.
All authors in Korea Astronomy and Space Science Institute (KASI) are the members of Dedicated Researchers for Extragalactic AstronoMy (DREAM). M.~Kim and S.-C.~Yang were supported by KASI-Carnegie Fellowship Program jointly managed by KASI and the Observatories of the Carnegie Institution for Science. This work was supported by the National Research Foundation of Korea (NRF) grant, No. 2008-0060544, funded by the Korea government (MSIP).

\begin{deluxetable}{rllcr @{$\pm$} lr @{.} lr @{.} l}
\tablenum{2} \tablecolumns{10} \tablecaption{ Results of Linear Regression, Correlation Analysis and Permutation Test\label{rand}} \tablewidth{0pt}
\tablehead{\multicolumn{2}{c}{Figure} & Condition & Sample $^{(i)}$ & \multicolumn{2}{c}{Slope $^{(ii)}$} & \multicolumn{2}{c}{Correlation} & \multicolumn{2}{c}{Significance Level $^{(iii)}$} \\
& & & Size & \multicolumn{2}{c}{ } & \multicolumn{2}{c}{Coefficient} & \multicolumn{2}{c}{of Null Hypothesis} }
\startdata
\multicolumn{10}{l}{(Cluster-centric distance)} \\
\ref{cldist}
& (a) & $N_{\textrm{\scriptsize com}}\ge1$ & 60 & $-0.088$ & 0.130 & $-0$ & 104 & 45 & $3\%$ \\
& (b) & $N_{\textrm{\scriptsize com}}\ge2$ & 45 & $0.031$ & 0.074 & $0$ & 047 & 78 & $2\%$ \\
& (c) & $N_{\textrm{\scriptsize com}}\ge1$, $+$ photo-z & 65 & $-0.101$ & 0.124 & $-0$ & 120 & 34 & $1\%$ \\
& (d) & $N_{\textrm{\scriptsize com}}\ge2$, $+$ photo-z & 50 & $-0.003$ & 0.082 & $-0$ & 004 & 97 & $8\%$ \\
& (e) & $N_{\textrm{\scriptsize com}}\ge1$, $-$ photo-z & 52 & $-0.143$ & 0.151 & $-0$ & 158 & 27 & $3\%$ \\
& (f) & $N_{\textrm{\scriptsize com}}\ge2$, $-$ photo-z & 37 & $0.001$ & 0.087 & $0$ & 001 & 99 & $3\%$ \\
\hline
\multicolumn{10}{l}{(Local luminosity density)} \\
\ref{lden}
& (a) & $N_{\textrm{\scriptsize com}}\ge1$ & 60 & $0.030$ & 0.039 & $0$ & 100 & 52 & $2\%$ \\
& (b) & $N_{\textrm{\scriptsize com}}\ge2$ & 45 & $0.011$ & 0.030 & $0$ & 048 & 77 & $9\%$ \\
& (c) & $N_{\textrm{\scriptsize com}}\ge1$, $+$ photo-z & 65 & $0.037$ & 0.040 & $0$ & 120 & 37 & $3\%$ \\
& (d) & $N_{\textrm{\scriptsize com}}\ge2$, $+$ photo-z & 50 & $0.021$ & 0.032 & $0$ & 084 & 57 & $5\%$ \\
& (e) & $N_{\textrm{\scriptsize com}}\ge1$, $-$ photo-z & 52 & $0.044$ & 0.042 & $0$ & 145 & 28 & $0\%$ \\
& (f) & $N_{\textrm{\scriptsize com}}\ge2$, $-$ photo-z & 37 & $0.027$ & 0.033 & $0$ & 116 & 46 & $6\%$ \\
\hline
\multicolumn{10}{l}{(Bright galaxy $i$-band magnitude)} \\
\ref{hostmag}
& (a) & $N_{\textrm{\scriptsize com}}\ge1$ & 60 & $-0.001$ & 0.014 & $-0$ & 011 & 93 & $4\%$ \\
& (b) & $N_{\textrm{\scriptsize com}}\ge2$ & 45 & $-0.003$ & 0.014 & $-0$ & 029 & 86 & $0\%$ \\
& (c) & $N_{\textrm{\scriptsize com}}\ge1$, $+$ photo-z & 65 & $-0.001$ & 0.014 & $-0$ & 004 & 97 & $0\%$ \\
& (d) & $N_{\textrm{\scriptsize com}}\ge2$, $+$ photo-z & 50 & $-0.002$ & 0.014 & $-0$ & 016 & 91 & $1\%$ \\
& (e) & $N_{\textrm{\scriptsize com}}\ge1$, $-$ photo-z & 52 & $-0.004$ & 0.015 & $-0$ & 032 & 82 & $0\%$ \\
& (f) & $N_{\textrm{\scriptsize com}}\ge2$, $-$ photo-z & 37 & $-0.006$ & 0.014 & $-0$ & 064 & 71 & $6\%$ \\
\hline
\multicolumn{10}{l}{(Bright galaxy $i$-band light concentration)} \\
\ref{hostcon}
& (a) & $N_{\textrm{\scriptsize com}}\ge1$ & 60 & $-0.012$ & 0.038 & $-0$ & 034 & 83 & $6\%$ \\
& (b) & $N_{\textrm{\scriptsize com}}\ge2$ & 45 & $-0.030$ & 0.045 & $-0$ & 096 & 52 & $9\%$ \\
& (c) & $N_{\textrm{\scriptsize com}}\ge1$, $+$ photo-z & 65 & $-0.010$ & 0.037 & $-0$ & 026 & 83 & $7\%$ \\
& (d) & $N_{\textrm{\scriptsize com}}\ge2$, $+$ photo-z & 50 & $-0.024$ & 0.046 & $-0$ & 072 & 59 & $0\%$ \\
& (e) & $N_{\textrm{\scriptsize com}}\ge1$, $-$ photo-z & 52 & $-0.007$ & 0.040 & $-0$ & 021 & 89 & $8\%$ \\
& (f) & $N_{\textrm{\scriptsize com}}\ge2$, $-$ photo-z & 37 & $-0.025$ & 0.051 & $-0$ & 079 & 64 & $0\%$ \\
\hline
\multicolumn{10}{l}{(Bright galaxy $r-i$ color)} \\
\ref{hostcol}
& (a) & $N_{\textrm{\scriptsize com}}\ge1$ & 60 & $0.300$ & 0.177 & $0$ & 206 & 8 & $9\%$ \\
& (b) & $N_{\textrm{\scriptsize com}}\ge2$ & 45 & $0.360$ & 0.167 & $0$ & 375 & 1 & $3\%$ \\
& (c) & $N_{\textrm{\scriptsize com}}\ge1$, $+$ photo-z & 65 & $0.226$ & 0.161 & $0$ & 172 & 15 & $0\%$ \\
& (d) & $N_{\textrm{\scriptsize com}}\ge2$, $+$ photo-z & 50 & $0.242$ & 0.157 & $0$ & 271 & 6 & $3\%$ \\
& (e) & $N_{\textrm{\scriptsize com}}\ge1$, $-$ photo-z & 52 & $0.398$ & 0.183 & $0$ & 258 & 5 & $8\%$ \\
& (f) & $N_{\textrm{\scriptsize com}}\ge2$, $-$ photo-z & 37 & $0.461$ & 0.175 & $0$ & 464 & 0 & $7\%$ \\
\enddata
\tablecomments{$(i)$ The number of bright galaxies satisfying each condition.
$(ii)$ The error values are the Bootstrap uncertainties.
$(iii)$ Results from permutation tests. The smaller value implies the stronger correlation.}
\end{deluxetable}

\end{document}